\documentclass[smallextended]{svjour3}       
\usepackage[margin=0.7in]{geometry}
\usepackage{amsmath}
\usepackage{amsfonts}
\newcommand{\subparagraph}{}
\usepackage{titlesec}
\usepackage{tikz}
\usepackage{makecell}
\usepackage{subfig}
\usepackage[section]{placeins}
\usepackage[style=numeric-comp, maxbibnames=99, backend=biber, bibencoding=utf8]{biblatex}
\bibliography{refclean}
\usepackage{enumitem}

\newcommand{\textsharp}{$\sharp$}

\newcommand\MIREX{\textsc{mirex}}
\newcommand\MIR{\textsc{mir}}
\newcommand\HCI{\textsc{hci}}
\newcommand\PP{\textsc{pp}}
\newcommand\CC{\textsc{cc}}
\newcommand\PCA{\textsc{pca}}
\newcommand\LVQ{\textsc{lvq}}
\newcommand\LDA{\textsc{lda}}
\newcommand\SVM{\textsc{svm}}
\newcommand\NB{\textsc{nb}}

\newcommand\MIDI{\textsc{midi}}

\newcommand\FANTASTIC{\textsc{fantastic}}

\newcommand\TEC{\textsc{tec}}
\newcommand\MTP{\textsc{mtp}}

\newcommand\JKUPDD{\textsc{jku-pdd}}
\newcommand\MTCANN{\textsc{mtc-ann}}

\newcommand\SIA{\textsc{sia}}
\newcommand\SIATEC{\textsc{siatec}}
\newcommand\COSIATEC{\textsc{cosiatec}}
\newcommand\SIATECC{\textsc{siatecc}}
\newcommand\COSIATECC{\textsc{cosiatec/c}}
\newcommand\OL{\textsc{ol}}
\newcommand\OLL{\textsc{ol}\oldstylenums{1}}
\newcommand\OLLL{\textsc{ol}\oldstylenums{2}}
\newcommand\VM{\textsc{vm}}
\newcommand\VMM{\textsc{vm}\oldstylenums{1}}
\newcommand\VMMM{\textsc{vm}\oldstylenums{2}}
\newcommand\SIACF{\textsc{siacf}\oldstylenums{1}}
\newcommand\SIACR{\textsc{siacr}}
\newcommand\SIACP{\textsc{siacp}}
\newcommand\SIAR{\textsc{siar}}
\newcommand\SIARCT{\textsc{siarct-cfp}}
\newcommand\SIACFP{\textsc{siacfp}}
\newcommand\SYMCHM{\textsc{symchm}}
\newcommand\SC{\textsc{sc}}
\newcommand\ME{\textsc{me}}
\newcommand\PATMINR{\textsc{p}at\textsc{m}inr}
\newcommand\MGDP{\textsc{mgdp}}

\usetikzlibrary{shapes,arrows}

\begin{document}

\title{A Computational Evaluation of Musical Pattern Discovery Algorithms }

\author{Iris Yuping Ren, Anja Volk, Wouter Swierstra, Remco C. Veltkamp}
\institute{I. Ren \at
             Department of Information and Computing Sciences, Utrecht
             University\\
             \email{y.ren@uu.nl}
             \and
             A. Volk \at
             Department of Information and Computing Sciences, Utrecht University
             \and
             W. Swierstra \at
             Department of Information and Computing Sciences, Utrecht University
             \and
             R. C. Veltkamp \at
             Department of Information and Computing Sciences, Utrecht University
           }

\maketitle
           
 \begin{abstract}
     Pattern discovery algorithms in the music domain aim to find meaningful components in musical compositions. 
     Over the years, although many algorithms have been developed for pattern discovery in music data, it remains a challenging task.
     To gain more insight into the efficacy of these algorithms, we introduce three computational methods for examining their output: Pattern Polling, to combine the patterns; Comparative Classification, to differentiate the patterns; Synthetic Data, to inject predetermined patterns. 
     In combining and differentiating the patterns extracted by algorithms, we expose how they differ from the patterns annotated by humans as well as between algorithms themselves, with rhythmic features contributing the most to the algorithm-human and algorithm-algorithm discrepancies. 
     Despite the difficulty in reconciling and evaluating the divergent patterns extracted from algorithms, we identify some possibilities for addressing them. 
     In particular, we generate controllable synthesised data with predetermined patterns planted into random data, thereby leaving us better able to inspect, compare, validate, and select the algorithms.
     We provide a concrete example of synthesising data for understanding the algorithms and expand our discussion to the potential and limitations of such an approach. 

\keywords{Music structure \and Synthetic Data \and Classification \and Pattern discovery}
 \end{abstract}

\section{Introduction}\label{sec:intro}
\textbf{Patterns and algorithmic musical pattern discovery}
Patterns are ubiquitous in nature.
Finding patterns involves identifying regularities, such as repetitions and variations.
Structured repetitions and different variations are amongst the fundamental elements of music.
Algorithmic musical pattern discovery research aspires to uncover and extract such elements automatically. 
The task of discovering patterns in music is non-trivial as we would like to illustrate with examples below. 

\begin{figure}
    \begin{center}
    \includegraphics[width=\linewidth]{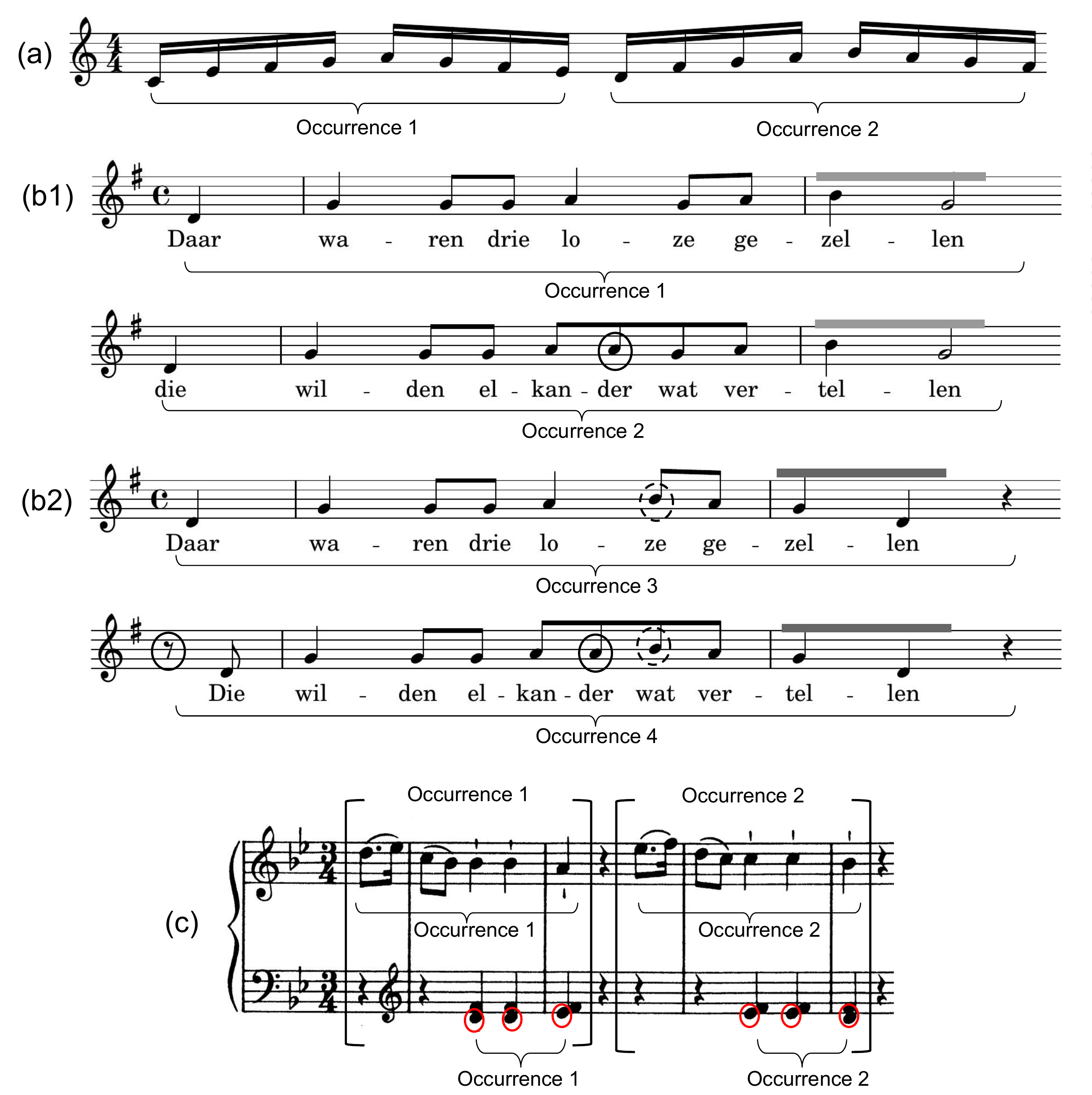}
  \end{center} 
  \caption{Example musical patterns in (a) Piano Étude (b1-2) Folk songs (c) Classical music.
    (a) The second occurrence is a transposition of the first occurrence higher in pitch. 
    (b1-2) b1 and b2 are two Dutch folk songs that belong to the same tune family. Each line (three bars) can be considered a pattern occurrence. The single note differences between occurrences within a song are circled with a solid outline; the differences across the two songs are circled with a dashed outline. Two notes differences between the two songs are marked with top bars.
    (c) On the upper staff, we see a similar tonal transposition as (a); at the bottom, we circle out the changed notes. These changes can also be described using the change of harmony.   }
  \label{fig:examples}
\end{figure}

\textbf{Examples}
A pattern normally has multiple occurrences in music, and collections of these occurrences constitute musical patterns. 
In Figure~\ref{fig:examples}, we show three examples, to give an intuitive and concrete idea of patterns that can be found in music.
Without relying on specific musical terms, we can describe how these pattern occurrences relate to each other as:
in (a), the second occurrence is based on the first but higher;
in (b1), the second line of melody is almost an exact repetition of the first line, but with a slight change in the middle;
in (b2), the entire melody is almost an exact repetition of (b1), but with slight changes at the beginning and the end. 
in (c), the two bracketed areas are largely the same but carry subtle differences.
With more precise vocabulary in music, we can describe them as:
in (a), the second occurrence is has the same shape as the first occurrence but transposed upward by a whole tone;
in (b1), a crotchet in the first line is split into two quivers in the second;
in (b2), the same rhythmic changes as in (b1) apply; in addition, the second line starts off-beat, and pitches of three notes are changed in comparison to (b1). 
in (c), the melodic motif of the first occurrence is transposed by a semitone in the second occurrence and the underlying harmony changes from B flat major to F dominant seventh, and then from F dominant seventh back to B flat major again.

\begin{figure}
    \begin{center}
  \includegraphics[width=0.5\linewidth]{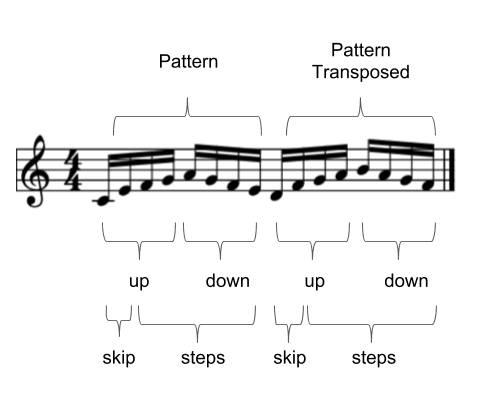}
\end{center}
\caption{Different patterns in piano étude.
  In addition to the patterns highlighted in Figure~\ref{fig:examples}, we can also identify the two other possibilities as bracketed under the stave as patterns: a combination of repeated upward/downward motions or different intervallic combinations. }
  \label{fig:examplehanon}
\end{figure}

\begin{figure}    
    \begin{center}
  \includegraphics[width=0.5\linewidth]{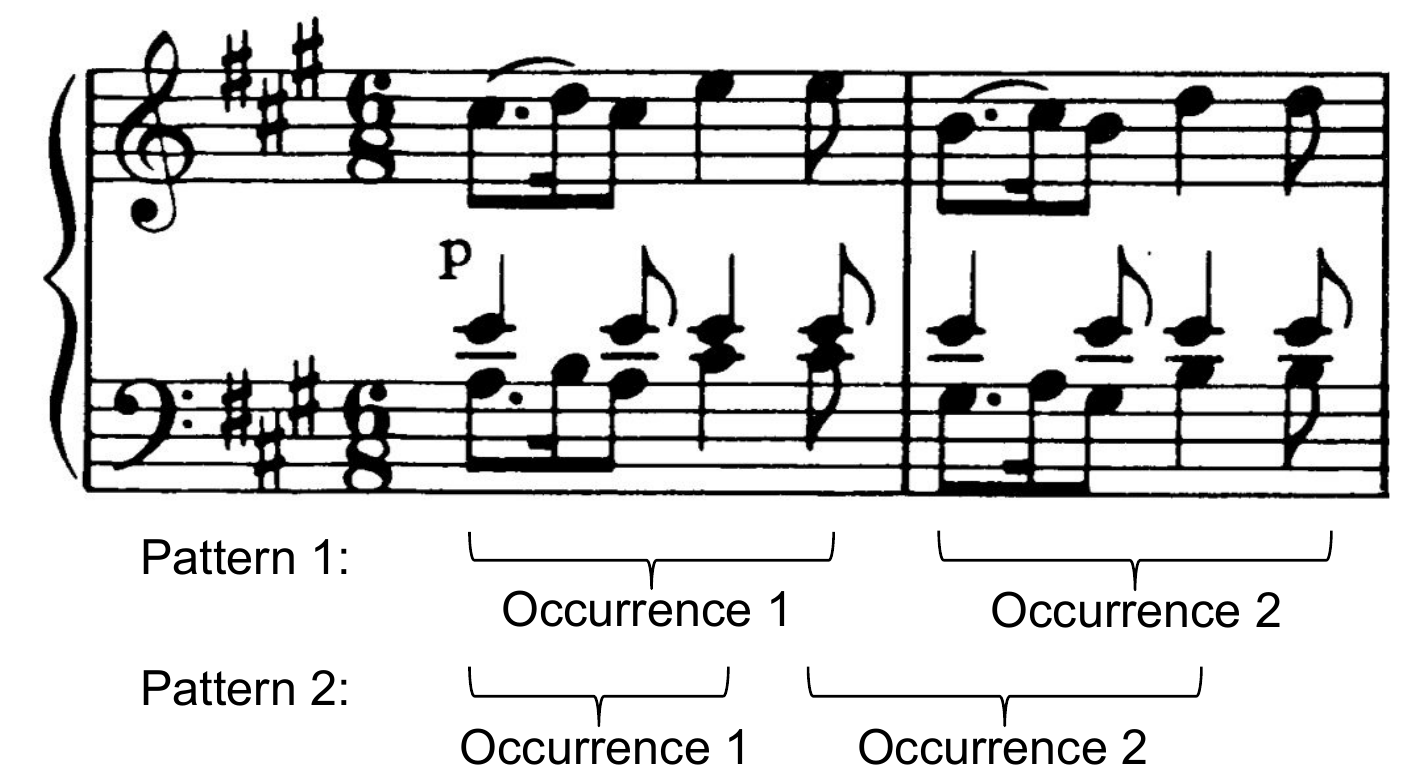}
\end{center}
\caption{Overlapping patterns in Mozart's Adagio K.331. This musical passage is ambiguous in terms of the musical patterns it contains. What are musical patterns, in this case, is open to interpretation and can be performed and perceived subjectively~\cite{lerdahl1985generative,gabrielsson1987once}. }
  \label{fig:examplemozart}
\end{figure}

The descriptions above give just one possible set of relationships between pattern occurrences amongst many.
In other words, given the same musical extracts, there might be alternative or multiple interpretations of the patterns contained within them.
For example, upon closer inspection of Figure~\ref{fig:examples} (a), we can also find hierarchical structure in patterns, such as those in Figure~\ref{fig:examplehanon}.
In Figure~\ref{fig:examplemozart}, the example shows how listeners can have two different interpretations of what the musical patterns are~\cite{lerdahl1985generative}. 
In fact, this example has been intensively studied: \cite{gabrielsson1987once} identified that this piece has been published in different editions with different phrasings, and piano performances of the piece with differing interpretations that correspond to these different phrasings were analysed. 
From these examples, we can see the diversity in musical patterns and therefore begin to appreciate the substantial complexity of automatic musical pattern discovery.
This complexity includes, but is not limited to, reconciling different time-scales, different types of patterns, and different degrees of subjectivity and ambiguity. 

\textbf{Research landscape and applications}
Research into musical patterns and pattern discovery algorithms is of broad and current relevance~\cite{janssen2013finding, janssenretained, collins2013discovery, ren2017search, ren153analysis, ren2016closed, meredith2002algorithms, lartillot2014patminr, pesek2017symchm, conklin2001representation, velarde2013approach, hsu1998efficient, velarde2016wavelet, meredith2013cosiatec, collins2013siarct, nieto2014identifying, collins2011improved, forth2012cognitively, conklin2011comparative, rolland1999discovering, meredith2016using}. 
With recent, rapid development in research areas such as pattern recognition and machine learning, many algorithms that are potentially suitable for automatically extracting musical patterns become available.
Together with the curiosity-driven pursuit of unveiling hidden structures in music data, algorithmic musical pattern discovery inspires many to explore and understand music from its modular components and their combinations.
Another motivation driving the development of musical pattern discovery is the
potential for applications in areas such as: genre classification, error correction, automatic transcription, and segmentation in Music Information Retrieval (\MIR{}, \cite{melendez2011unsupervised, cambouropoulos2006musical, conklin2006segmental, lin2004music, dixon2004towards, conklin2010discovery}); and automatic composition, digital humanities, and learning systems in Human-Computer Interaction (\HCI{}, \cite{collins2011improved, fowler1966discovery, kaplan2015map}).
For example, given that composers employ patterns to introduce structure into their music~\cite{hsu1998efficient}, the discovered patterns could be used to provide auto-completion and inspirational suggestions;
we can also highlight discovered patterns in music as a guide for attentive listening, memory anchoring, and efficient practising~\cite{kubik1979pattern,jones1987dynamic};
for musicologists and music theorists, facing the complexity of large corpora, the algorithmically discovered pattern candidates can provide support and evidence for categorisation and theorisation efforts~\cite{lerdahl1985generative,agawu2014music,gjerdingen2007music,huron2006sweet,zbikowski2002conceptualizing}.

\textbf{Motivation - Challenges}
In spite of the existence of many generic pattern discovery algorithms~\cite{parida2007pattern, brand1999pattern, Wang1994, bertens2016keeping,  cooley1997web} and musical pattern discovery algorithms~\cite{lartillot2014patminr, pesek2017symchm, conklin2001representation, velarde2013approach, hsu1998efficient, velarde2016wavelet, meredith2013cosiatec, collins2013siarct, nieto2014identifying, collins2011improved, forth2012cognitively, conklin2011comparative, rolland1999discovering, meredith2016using} (see~\cite{janssen2013finding} for a detailed overview), there have been persistent challenges in comparing and improving these algorithms.
The challenges range from methodological questions, such as the ground-truth controversy (whether automatically extracted patterns should conform to human annotations or can be meaningful though incomprehensible to humans) to the practical problems surrounding untangling multiple parameters and filters in the algorithms and the common undesirable outcome of a large number of extracted patterns. 
Even making the best effort to take the ground-truth ``out-of-the-loop''~\cite{boot2016evaluating}, we experience poor outcomes in compression and classification tasks when these automatically extracted patterns were employed.
Automating musical pattern discovery is not yet a solved problem.
This paper is an extension of our recent work on leveraging different visualisation methods and classification algorithms~\cite{ren2017search,ren153analysis} to show differences between human annotations and the output of state-of-the-art musical pattern discovery algorithms; using synthetic data is a new contribution of this paper.
Following a similar direction, this paper aims to analyse disagreement between pattern discovery algorithms and between algorithmically extracted patterns and human-annotated patterns.
Our intention is not to create metrics to rank the algorithms, but to introduce methods to examine their output, which could potentially inform us more about the algorithms as well as the music data being used.

\vspace{5mm}
\textbf{Contributions}
We propose three methods to inspect the output from musical pattern discovery algorithms, corresponding to the contributions and sections in this paper as follows:
\begin{itemize}
\item Pattern Polling (\PP{}): We devise PP to combine the output from different algorithms, following fusion and ensemble methods commonly used in improving pattern recognition accuracy. We show that there are promising correspondences between the combined results and human annotations, but that discrepancies between output from different algorithms constitute an obstacle to any significant improvement. (Section 3)
\item Comparative Classification (\CC{}): To verify the discrepancies between the algorithms, we leverage the discriminative power of classification algorithms to compare features extracted from algorithmically discovered, hand-annotated, and randomly selected patterns. We found that rhythmic features are the most important in differentiating between these pattern groups. (Section 4)
\item Synthetic data with planted patterns: To better understand the output from different algorithms, we examine them using synthetic data---random data with musical patterns planted within it. We found that some algorithms perform as expected while others do not. (Section 5)
\end{itemize}

In this work, with a focus on symbolic monophonic melodies, we provide a comparison and evaluation of musical pattern discovery algorithms using these three methods.

\section{Previous work: musical pattern discovery algorithms}\label{sec:alg}
In this section, we provide more information for previous work.  
In particular, we introduce musical pattern discovery algorithms that we use for our experiments in the next sections.
Each algorithm is accompanied by a short description. 

\textbf{MIREX (Music Information Retrieval Evaluation eXchange)}
In the field of \MIR{}, \MIREX{} is the evaluation platform \cite{mirex} where tasks related to information retrieval in music data are organised, such as the task of structural segmentation and symbolic melodic similarity. 
In relation to musical pattern discovery, the Discovery of Repeated Themes \& Sections task \cite{collins2013discovery} has received participation and submissions from musical pattern discovery researchers.
Variants of standard evaluation metrics--precision, recall, and F1 score--are used in this task~\cite{collins2013discovery}.
Other evaluation endeavours have used a variety of datasets and metrics \cite{meredith2002algorithms,conklin2001representation,conklin2008feature,forth2009approache,knopke2009system,dannenberg2003pattern,weiss2010identifying}.
Most evaluation rank the algorithms according to a score as the proxy for how meaningful the discovered patterns are.
However, these metrics are not entirely without controversy: the performance of algorithms varies on different metrics and different musical pieces, for example, which makes it difficult to analyse the overall and specific strengths and limitations of algorithms.
Problems such as these motivate us to examine the output patterns from the algorithms in more detail to advance this area of research.
Based on availability and compatibility of the algorithms, we examine algorithms as listed in Table \ref{tab:algs}, consisting of a range of algorithms submitted to \MIREX{} as well as other general musical pattern discovery algorithms. 
In the following paragraphs, we provide short descriptions of the algorithms we
examine in Section \ref{sec:ismir}, \ref{sec:class}, and \ref{sec:synth},

\textbf{Wavelet-based}\cite{velarde2014wavelet}
Wavelets and related applications have a strong theoretical foundation in mathematics and physics. 
By convolving wavelets with symbolic melodies, we obtain correlations between the unknown symbolic melodies and the known patterns encoded in the wavelets.
The resulting correlations are used for melodic segmentation and concatenation, which is then compared, clustered, and ranked.
These ranked clusters then become the melodic patterns retrieved from the piece.
The algorithm has two \textsc{matlab} implementations using values for the clustering threshold parameter.

\textbf{Compositional Hierarchical Model}\cite{pesek2017symchm}
This model was inspired by object categorisation in computer vision.
As an alternative to deep learning models, the model imposes a hierarchical structure in different compositional layers and tries to provide a transparent deep architecture to discover repetitions in music.
The method is based on the assumption that repetitive patterns can be characterised by the number of occurrences of their sub-patterns.
The model learns pitch patterns from symbolic data in an unsupervised manner. 

\textbf{MGDP (Maximally General Distinctive Pattern)}\cite{van2016pattern}
This algorithm computes the maximally general distinctive patterns in a corpus.
The algorithm is a sequential pattern mining method and works by constructing a suffix tree and performing statistical modelling. 

\textbf{PatMinr}\cite{lartillot2016automated}
The \PATMINR{} algorithm considers both the closed pattern and cyclic pattern criteria.
It uses an incremental one-pass approach to identify pattern occurrences, with consideration to various music attributes such as chromatic vs. diatonic pitch, metrical position, and articulation. 
A pattern prefix tree is employed in the implementation.
The algorithm is available in \textsc{matlab} as part of \MIR{}toolbox~\cite{lartillot2011mirtoolbox}. 

\textbf{MotivesExtractor}\cite{nieto2012perceptual}
MotivesExtractor employs perceptual grouping principles.
It computes a distance matrix between pairs of potential motives and then uses a clustering algorithm to identify the instances of the same pattern.
The implementation of this algorithm is in Python.

\textbf{Geometric Methods}
There exists a family of geometrically-inspired pattern discovery algorithms that can identify note groups with similar shapes and therefore extract these shapes as musical patterns.
Six algorithms, \SIA{}, \SIATEC{}, \COSIATECC{}ompress, \SIAR{}, \SIARCT{}, and Forth, all belong to this family of geometric musical pattern discovery algorithms. 
A tool written in Java is available online, supporting this array of point-set compression algorithms, including several \COSIATEC{} and \COSIATECC{}ompress approaches as well as Forth’s algorithm.

\textbf{SIA}\cite{meredith2002algorithms}
The algorithm computes vectors between one note and another and then takes the maximally repeated patterns in a multidimensional dataset. 
This method of pattern extraction can handle polyphony elegantly. 

\textbf{SIATEC}\cite{meredith2001pattern}
Based on the \SIA{} algorithm, \SIATEC{} generates a set of translational equivalence classes (\TEC{}s).
The algorithm computes the occurrences of all the maximally repeated patterns discovered by \SIA{}.

\textbf{COSIATEC/Compress}\cite{meredith2013cosiatec}
Both \COSIATEC{} and \SIATECC{}ompress iteratively apply the \SIATEC{} algorithm to compress a piece into the corresponding union of \TEC{}s of maximal translatable patterns (\MTP{}s).
The difference is that \COSIATEC{} finds the best \TEC{} and then removes its cover set from the input, while \SIATECC{}ompress extracts a list of \MTP\TEC{}s and then selects a subset of the best \TEC{}s to cover the input.
The selection criteria are based on compression ratio and compactness measures.  

\textbf{SIAR}\cite{collins2011improved}
Based on the \SIA{} algorithm, \SIAR{} limits the creation of vectors from all possible note combinations by constraining them to maximum of $R \in \mathbb{N}$ successive notes.
In this way, we might think of $R$ as the ``memory size'' parameter specifying the number of notes to consider in \SIA{}. 

\textbf{SIARCT-CFP}\cite{collins2013siarct}
This algorithm combines \SIA{} with a \textit{compactness trawler}, fingerprinting, and categorisation steps.
A compactness trawler removes all patterns that do not satisfy a parameterised compactness ratio $c \in \mathbb{R}$---the ratio between the number of notes in a pattern and the width of the bounding box of the pattern.
The fingerprinting step computes the rhythmic ratios in order to allow rhythmic variations.
The categorisation step ranks the discovered patterns in order of importance.
\SIARCT{} algorithm has an implementation in \textsc{matlab}.

\textbf{Forth}\cite{forth2012cognitively}
Based on the geometric method \SIATEC{}, the Forth algorithm incorporates considerations based on music theory and cognition.
It uses heuristic measures with compression ratio and compactness value thresholds.

\begin{table}
\begin{center}
\begin{tabular}{lcccc}
Algorithm & MIREX & PP & CC & Synth\\
\hline
\VMM{} (\textsc{w}avelet-Based) & x & x & x & x\\
\VMMM{} (\textsc{w}avelet-Based) & x & x & x & x\\
\SIACF{} (\SIATECC{}ompress) & x & x & x & x\\
\SIACR{} (\SIATECC{}ompress) & x & x & x & x\\
\SIACP{} (\SIATECC{}ompress) & x & x & x & x\\
\SC{} (\textsc{c}ompositional) & x & x & x & \\
\OLL{} (\PATMINR{}) & x & x &  & \\
\OLLL{} (\PATMINR{}) & x & x &  & \\
\ME{} & x & x &  & \\
\SIARCT{} & x & x &  & x\\
\MGDP{} &  &  &  & x\\
\COSIATEC{} &  &  &  & x\\
Forth &  &  &  & x\\
\end{tabular}
\end{center}
\caption{Algorithms used in Section \ref{sec:ismir} (PP and CC) and Section
  \ref{sec:synth} (Synth). A cross mark indicates that the algorithm is present. }
\label{tab:algs}
\end{table}

\textbf{Correspondence to the other sections}
In Table~\ref{tab:algs}, we present a summary of the algorithms and methods we use in Section~\ref{sec:ismir} and Section~\ref{sec:synth}.
The full name of the algorithms will be introduced in Section~\ref{subsec:visloc}.
There are algorithms that are not present for all experiments because of format compatibility.
This does not undermine the validity of our results because we use the algorithms individually and provide our analysis mostly on a case-by-case basis.
That is, we will look at algorithms individually most of the time, except in Section~\ref{subsec:combi}.
We expect that our analysis and methods could still be applied in the presence or absence of any number of other algorithms.

\begin{figure}[hbt!]
   \centerline{
 \includegraphics[width=0.8\columnwidth]{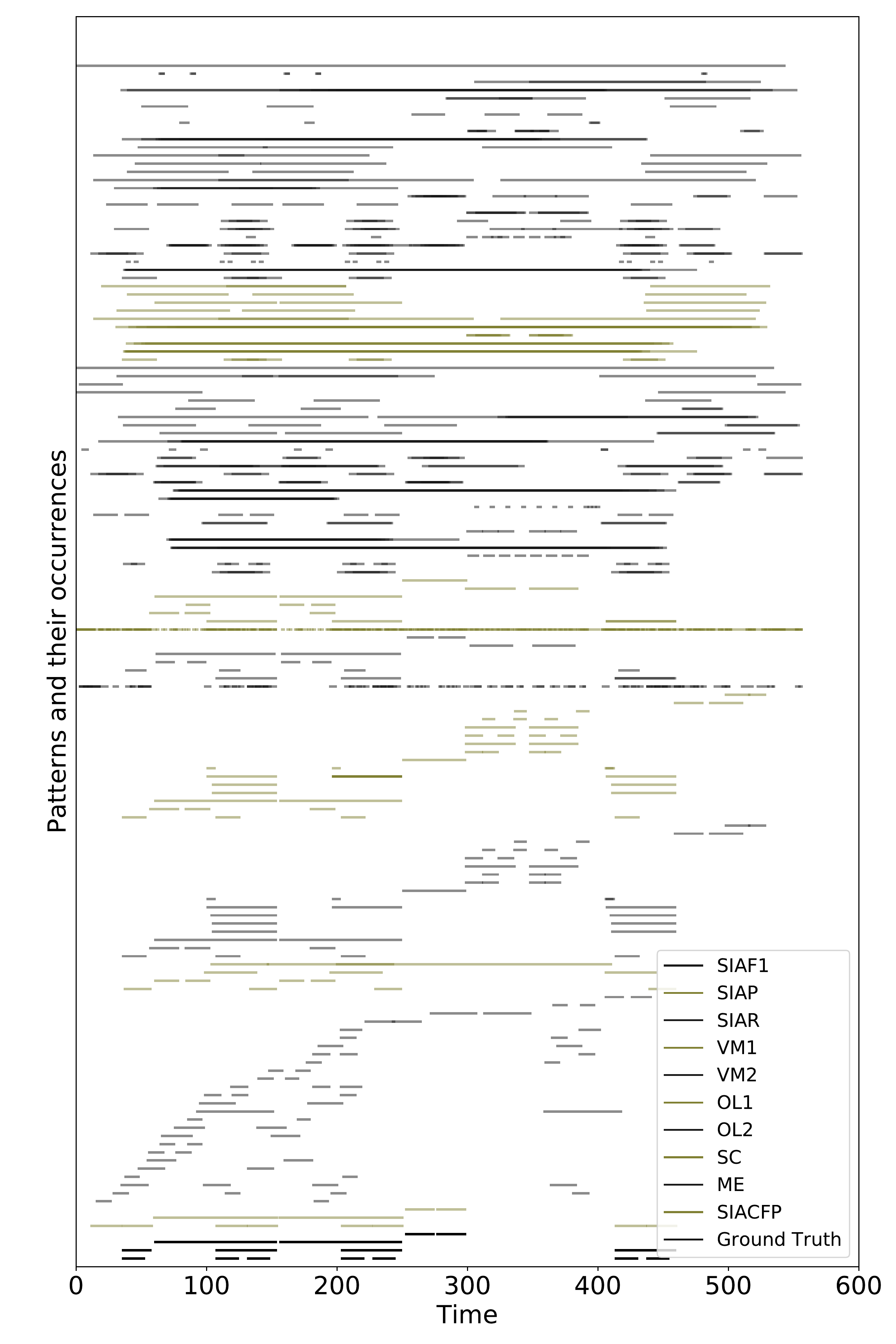}}
 \caption{Patterns extracted by all algorithms submitted to the \textsc{mirex}
   task 2014-2016 plus \textsc{siarct-cfp} on the monophonic version of Chopin's
   Mazurka Op.~24 No.~4. A horizontal bar shows the presence of a pattern. The
   x-axis represents time offset in crotchet units. The y-axis is not ordinal
   but merely separates individual pattern occurrences. The legend has the same order as the bars in the alternating colours, ground-truth being at the bottom. We can see the algorithms find different numbers of patterns and patterns of different lengths. }
 \label{fig:visual}
\end{figure}

\section{Combining and comparing the output of pattern discovery algorithms} \label{sec:ismir}
\textbf{Visualisation, combination, comparison, and analysis of algorithm output}
As motivated in Section~\ref{sec:intro}, we now describe the two experiments in which we combine and compare musical pattern discovery algorithms.
We visualise and inspect the algorithmically extracted musical patterns using their location in the musical pieces (first experiment) and their actual musical content (second experiment).
We also use two computational methods to further combine and compare the algorithms.
The computational method in the first experiment is inspired by data fusion and the second by classification.

\subsection{Visualising pattern location} \label{subsec:visloc}
Given a musical piece, the musical events in the musical pattern it contains are uniquely identified by the temporal locations of the pattern occurrences, which itself is uniquely identified by two values, the beginning and ending points.
Using these two values, we can concisely encode and represent musical pattern occurrences within a musical piece.
In Figure~\ref{fig:visual}, we visualise and emphasise the (non)existence of patterns by plotting a bar in between the beginning and ending points of the pattern occurrences. 

\textbf{Algorithm and musical pattern data}
We use the algorithms submitted to the \MIREX{} Discovery of Repeated Themes \& Sections task during 2014-2016: MotivesExtractor (\ME{})~\cite{nieto2014identifying}, \SIATECC{}ompress-\SIACP{} (\SIACP{}), \SIATECC{}ompress-\SIACF{} (\SIACF{}), \SIATECC{}ompress-\SIACR{} (\SIACR{})~\cite{meredith2016using}, \OLL{} \& \OLLL{}~\cite{lartillot2014patminr}, \VMM{} \& \VMMM{}~\cite{velarde2014wavelet}, \SYMCHM{} (\SC{})~\cite{pesek2017symchm}, and \SIARCT{} (\SIACFP{})~\cite{collins2013siarct}.
We take the monophonic version of Chopin's Mazurka Op.~24 No.~4 in the \JKUPDD{} dataset as our primary example case as shown in Figure~\ref{fig:visual}.
The \JKUPDD{} dataset~\cite{collins2013discovery} contains both symbolic and audio (synthesised from the score) data and human-annotated patterns of five musical pieces from Bach, Mozart, Beethoven, Chopin, and Gibbons.
The annotations are constructed from three sources~\cite{bruhn1993js, schoenberg1967fundamentals, barlow1948dictionary}.
A sample of the locations of the human annotations is shown in Figure~\ref{fig:examples}(c).

\textbf{Observations from the visualisation}
The output patterns of these state-of-the-art algorithms for our example piece are shown in Figure~\ref{fig:visual}.
We make several observations:
\begin{itemize}[noitemsep]
    \item Different algorithms find very different patterns---some tend to find shorter patterns, some longer; some find many patterns while others are more discerning.
    \item We have three algorithm families (\SIA{}, \VM{}, and \OL{}) which each consists of more than one algorithm. The algorithms from the same algorithm family tend to find similar patterns. Similarities observed include the number of patterns discovered, coverage of the song, and occurrence overlaps.
    \item The ground truth (which we take to be human annotations) is sparse in comparison to the patterns discovered by the algorithms.
    \item Taking an overview of the entire visualisation, we can see some correspondence and similarities between the algorithms and the ground truth patterns.
\end{itemize}
These observations hint at the possibility of combining similar results to devise an ensemble method as detailed below.

\subsection{Challenges of combining the output} \label{subsec:combi}
\textbf{Motivation}
Integrating different algorithms using data fusion has been shown to be a successful approach to improving overall performance in other areas dealing with ambiguous musical data, such as in automatic chord estimation~\cite{koops2016integration}.
For musical pattern discovery, there is hope to find a consensus between various algorithms to achieve an overall better pattern discovery result.
To this end, we devise the process of Pattern Polling (\PP{}), which takes the locations of discovered patterns from different algorithms as input, and outputs new locations of patterns based on this. 
We name the algorithm after ``Polling'' because it involves regularly measuring the level of consensus from an ensemble of algorithms, which is distinct from a random ``Sampling'' approach.

\textbf{PP design}
We develop the \PP{} algorithm based on the assumption that all pattern discovery algorithms aim to find passages containing a shared characteristic of interest---the ``patterns'' in musical compositions. 
We consider whether or not an algorithm recognises a pattern at any given time
point to count as a ``vote'' on whether that time point participates in this is
part of a musical pattern. 
For the sake of convenience, we define the salience degree of a time point as the number of discovered patterns at this time offset. 
In essence, \PP{} is a process in which each algorithm contributes to the salience degree of a time point based on their discovered patterns. 
The resulting \textit{polling curve} is then taken as a base to detect new patterns with input from all algorithms.

\begin{figure}      
  \centering
  \subfloat[Polling curve]{{\includegraphics[width=0.45\columnwidth]{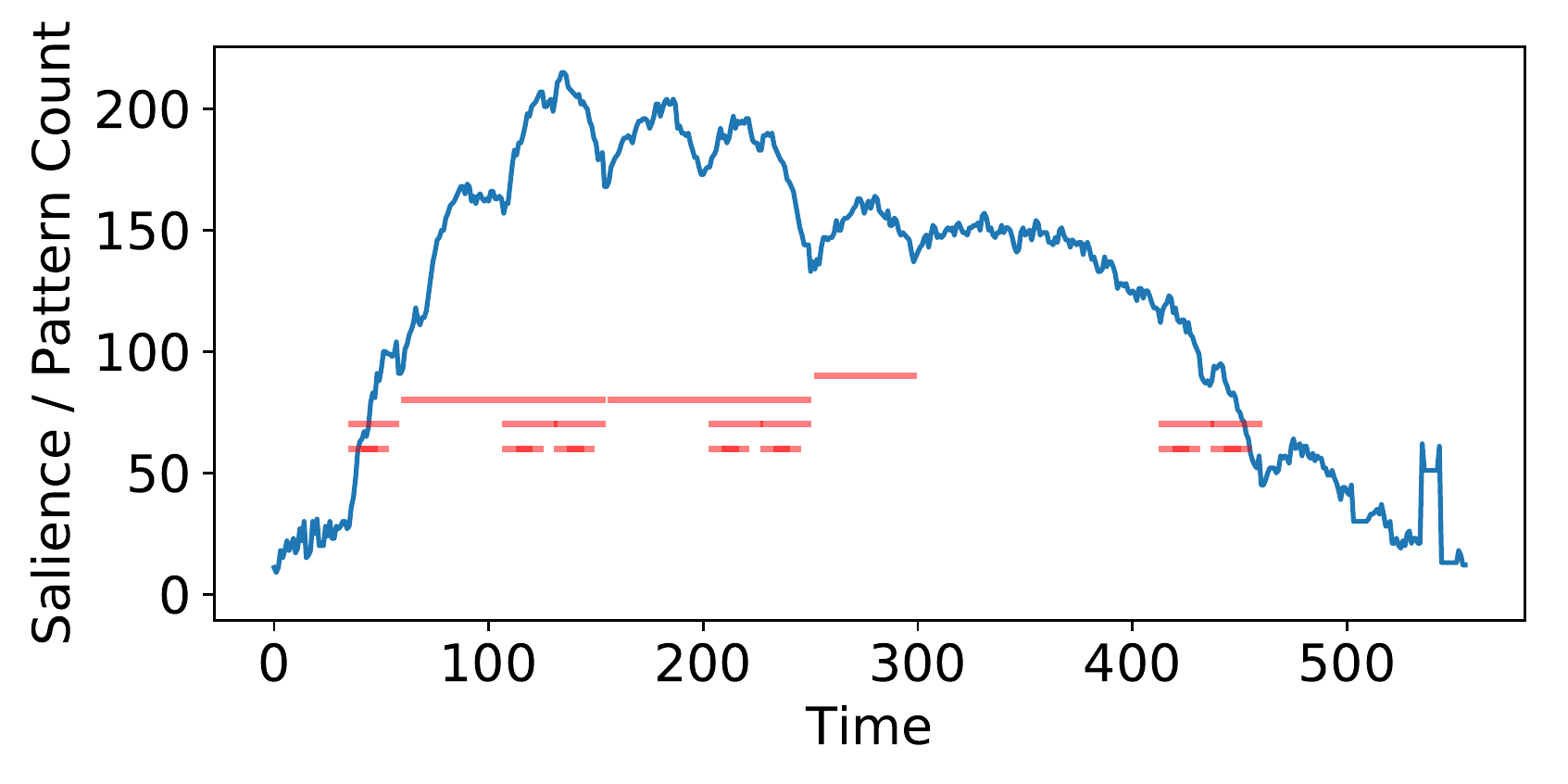} }}%
    \qquad
  \subfloat[Boundaries extracted using the \PP{} algorithm]{{\includegraphics[width=0.45\columnwidth]{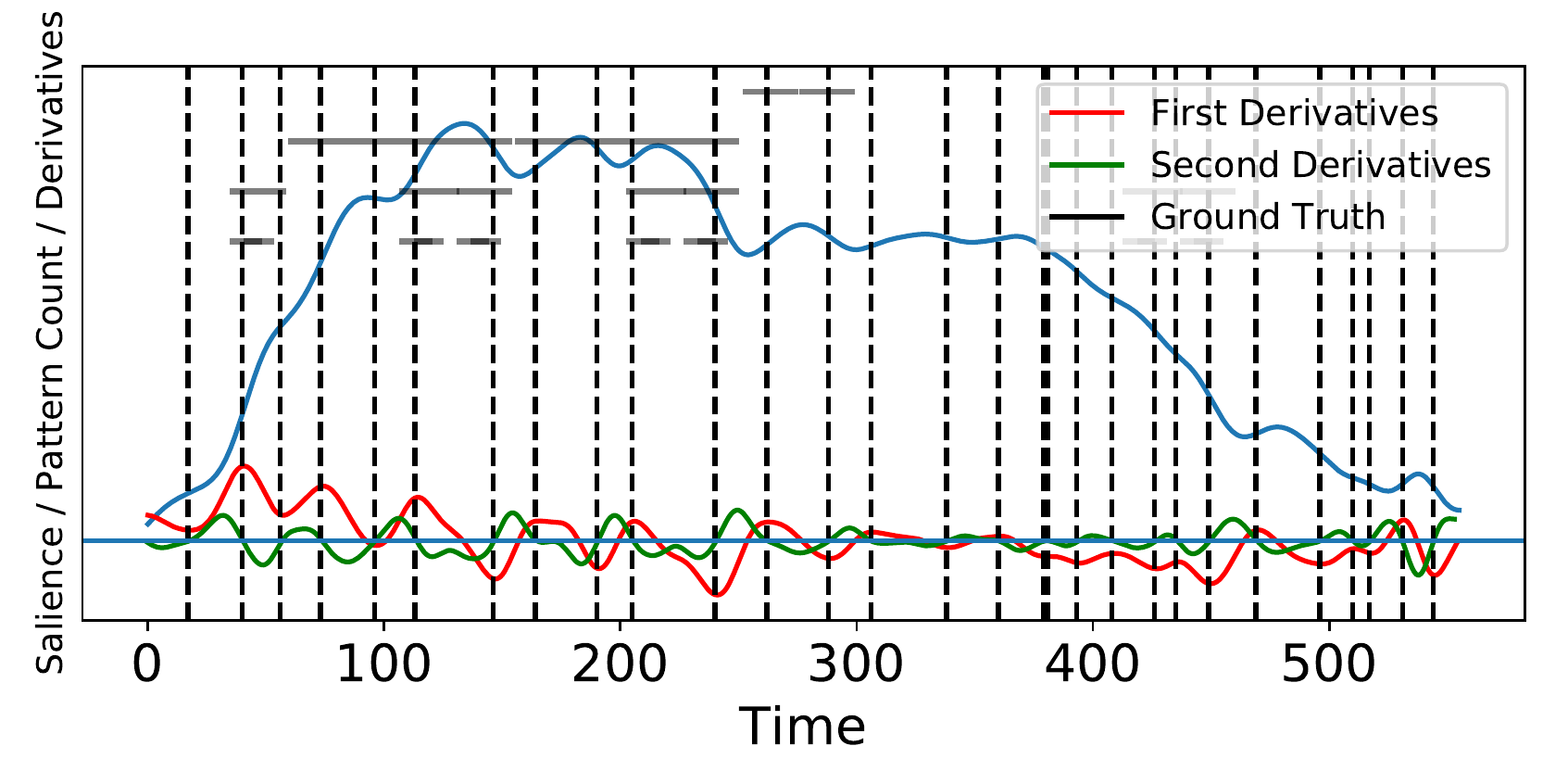} }}
 \caption{a. The polling curve of Chopin's Mazurka Op.~24 No.~4 using algorithms from the \MIREX{} task. 
 The horizontal bars show where the human-annotated patterns are present. 
 The x-axis represents time offset in crotchet units and the y-axis represents the salience value. 
 We see promising correspondences between the polling curve and human annotations. 
 b. Extracted pattern boundaries using \PP{} indicated with dashed vertical lines. 
 Many dashed lines are aligned with the boundaries of human annotations. 
 We also plotted the polling curve, the ground truth human annotations, as well as both the first and second derivatives for reference. }
 \label{fig:ppa}
\end{figure}

\textbf{Polling Curve}
An example of a polling curve using several algorithms from the \MIREX{} task is shown in Figure~\ref{fig:ppa} (a). 
To elaborate on how we calculate the polling curve, we start with discretised time points $T=[0, 1, ..., n]$ in the musical piece, with a resolution of one crotchet. 
If an algorithm finds a pattern occurrence at a given time point, we count that as one vote contributing to the pattern salience score at that time offset. 
We perform the same procedure for all pattern occurrences and sum the votes from all algorithms.
In the end, we obtain the polling curve $P(t)$, which is a time series consisting of the voting counts at each time offset $t \in T$. 

Mathematically, we have:
\begin{equation}\label{polling}
P(t) = \sum_{A} \sum_{P} \sum_{O} I^{A,P}_O(t)
\end{equation}
Where $A$ is the set of algorithms, $P$ is the set of patterns, $O$ is the set of occurrences, and $I_O^{a, p}(t)$ is the weighted indicator function of an occurrence of a pattern $p \in P$ in the output of an algorithm $a \in A$:
\begin{equation}\label{indicatoromega}
I^{a,p}_O(t)= 
\begin{cases}
\omega_a  &\text{$t \in o \subseteq p \subseteq a$}\\
0 &\text{$t \not\in o \subseteq p \subseteq a$}
\end{cases}
\end{equation}
Where $\omega_a$ is the weight assigned to an algorithm $a$. 

\textbf{What polling curves represent and how to extract patterns from the curve}
The polling curve uses the output from all individual pattern discovery algorithms. 
Given a music piece and a set of algorithms, it provides an estimate of how likely a region in music is to contain a pattern by taking into account all algorithmic output. 
The curve can be used for estimating where a pattern is more likely to be present in the musical piece.
In Figure~\ref{fig:ppa}, we see a promising level of agreement between this estimated likelihood and human-annotated patterns. 

To extract concrete patterns, we need to identify the beginnings and endings of patterns from the curve. 
The critical points of the curve can be helpful in this respect. 
Mathematically defined as the points at which the derivative of the curve is equal to zero, critical points indicate the prominent changes in the shape of curves. 
Although the polling curve is a discretised curve, we can compute the discrete derivatives and corresponding critical points.
These prominent shape changes then can be regarded as pattern boundaries.
To distinguish between the prominent changes and those too small to be relevant, we first perform a smoothing step on the polling curve. 

\textbf{Smoothing}\label{subsec:smoothing}
In our algorithm, we use the Savitzky-Golay filter~\cite{schafer2011savitzky} for smoothing, which is a linear least-squares polynomial fitting filter. 
When we apply smoothing, we reduce the impact of small irrelevant changes in the curve at the cost of potentially losing valuable details. 
With different degrees of smoothing, we capture different levels of detail in the polling curve. 
With this in mind, we make our \PP{} algorithm parametric on the degree of smoothness $s$.

\textbf{Derivative}\label{subsubsec:deri}
After smoothing, to find the prominent changes in the curve, we calculate the first and second discrete derivatives and take the critical points. 
More formally: let $P'(t) = P(t+1)-P(t)$ and let $P''(t) = P'(t+1)-P'(t)$, $t > 0, t \in T$.
We are interested in zero-crossing points ${\bar{t}}$ in $P'(t)$ and $P''(t)$ because each zero-crossing point $\bar{t}$ represents a change of direction in the polling curve. 
For example, when $P'(t) < 0$ and $P'(t+1) > 0$, we have a dipping point $P'(\bar{t}) = 0$ on the curve. 
There are more patterns discovered by the algorithms starting from this point, it is therefore likely to be a beginning of a pattern. 

One question remains as to how strong the dipping, tipping, concave, and convex positions in the curve should be if we are to pick them as boundaries. 
Here we introduce a second parameter of the \PP{} algorithm: a threshold on the steepness of the zero-crossing points $\lambda$. 
With different values of $\lambda$, we create a set of boundary sets which consist of time offsets at which zero-crossings occur. 

In Figure~\ref{fig:ppa}, we show an example of the extracted boundaries. 
We notice that some boundaries line up well with human-annotated pattern boundaries. 

\textbf{Summary: PP algorithm}
\PP{} starts by calculating a polling curve by taking into account the output of all algorithms. 
After smoothing the polling curve, the algorithm takes the critical points (i.e. the points at which the derivatives are equal to zero) of the curve as pattern boundaries. 

\textbf{Evaluation metrics}
To evaluate the accuracy of the extracted pattern boundaries, we choose to use the ground truth of human-annotated patterns.
We compare the computed boundaries with the beginnings and endings of patterns marked by humans with standard evaluation metrics of precision, recall, and F1 score. 
Following these standard evaluation metrics, we calculate the precision, recall, and F1 score with a degree of fuzziness, i.e. we attempt to match the boundaries with a tolerance of one crotchet note length as this is the degree of discretisation we used for creating the polling curve. 

\textbf{Results}
Using the setup above, we extract patterns using the \PP{} algorithm in a subset of \JKUPDD{}. 
The original \JKUPDD{} dataset contains five pieces in both polyphonic and monophonic format. 
We take three pieces in the monophonic format: Chopin's Mazurka Op.~24 No.~4, Mozart's Piano Sonata K.~282, 2nd movement, and Beethoven Piano Sonata Op.~2 No.~1, 3rd movement. 
The other two pieces contain a concatenation of voices from the polyphonic version, which violates the assumptions of certain algorithms, and are therefore excluded. 
 
The results are shown in Table~\ref{tab:var1}. 
By calculating the mean results of all algorithms, we can see that the best F1 score of \PP{} $0.360$ is better than the mean of the F1 scores of individual algorithms $0.3549$. 
When we look at the individual algorithms, the best F1 score of \PP{} ranks fifth out of ten. 
The \SIACFP{} algorithm performs the best overall. 
Although we also observe that \PP{} performs slightly better than the average of the individual algorithms, we cannot yet conclude that this fusion method improves the accuracy of pattern discovery significantly. 

\begin{table}
\small
 \begin{center}
 \begin{tabular}{c|c|c|c}
  Algorithm & Precision & Recall & F1 \\
\hline
\textsc{me} & (0.125, 0.086) & (0.184, 0.077) & (0.149, 0.083)\\
\textsc{sc} & (0.396, 0.022) & (0.419, 0.068) & (0.402, 0.046)\\
\textsc{ol}\oldstylenums{1} & (0.420, 0.038) & (0.565, 0.044) & (0.462, 0.023)\\
\textsc{ol}\oldstylenums{2} & (0.422, 0.061) & (0.565, 0.044) & (0.483, 0.054)\\
\textsc{siaf}\oldstylenums{1} & (0.139, 0.049) & (0.670, 0.005) & (0.228, 0.041)\\
\textsc{siar} & (0.213, 0.039) & (0.427, 0.000) & (0.279, 0.021)\\
\textsc{siap} & (0.117, 0.043) & (0.596, 0.008) & (0.195, 0.037)\\
\textsc{vm}\oldstylenums{1} & (0.137, 0.035) & (\textbf{1.0}, \textbf{0.0}) & (0.240, 0.029)\\
\textsc{vm}\oldstylenums{2} & (0.206, 0.073) & (0.543, 0.024) & (0.296, 0.060)\\
\textsc{siacfp} & (\textbf{0.819}, 0.030) & (0.82, 0.064) & (\textbf{0.815}, 0.046)\\
\hline
\textsc{pp-p} & \textbf{0.478} & 0.206 & 0.249\\
\textsc{pp-r}  & 0.228 & \textbf{0.867} & 0.35\\
\textsc{pp-f}\oldstylenums{1} & 0.248 & 0.738 & \textbf{0.360}\\
 \end{tabular}
\end{center}
 \caption{\MIREX{}: (Mean, Variance) of the precision, recall, and F1 score of the pattern boundaries of different algorithms. 
 The \textsc{pp-p}, \textsc{pp-r} and \textsc{pp-f}\oldstylenums{1} are obtained using a 3-fold cross-validation training process optimising precision, recall, and F1 score. 
 Because we only have one piece in the test set, there is no variance value. 
 The best results from individual algorithms and \textsc{pp} are shown in bold. }
 \label{tab:var1}
\end{table}

\textbf{Why is thereby no significant improvement?}
From the results, we identify some potential reasons as to why \PP{} does not outperform individual algorithms.
Firstly, the available dataset is small and the human-annotated patterns are sparse, which is problematic for training the parameters in \PP{}. Secondly, the algorithms disagree with each other on pattern length, pattern overlap, and the number of patterns, which may be caused by a variety of different factors including: the inherent ambiguity of music and pattern perception; the lack of a unified goal; the different target applications of the musical pattern discovery algorithms; or a combination of all of these factors.
In the end, although we observed a promising level of consensus between algorithms in Figure~\ref{fig:visual} and Figure~\ref{fig:ppa}, Table~\ref{tab:var1} reveals that this is not yet sufficient to extract patterns that substantially agree with the human-annotated patterns.

\textbf{The use of human annotations as ground truth}
Whether or not algorithms should be evaluated based on the extent to which their output reflects human annotations is case-dependent, given that human annotators are known to disagree amongst themselves.
For this reason, aspects of subjectivity should be taken into account when comparing algorithmic output with human annotations. 
In addition, as we can see that pattern discovery algorithms produce very different output, one might ask whether or not the output of different algorithms might be useful for different application scenarios.
To give a hypothetical example, one algorithm could be more suitable for extracting patterns for educational purposes and another could be more suitable for improving the segmentation task in MIR.
There may be overlaps amongst different application scenarios where the output patterns can be used for multiple purposes.
However, there is a lack of an agreed-upon approach to categorising and formalising different pattern extraction algorithms according to their potential applications.
In the second experiment, we explore a potential starting point towards realising such an approach by looking into the properties of patterns extracted by different algorithms. 

\textbf{From location to content, from combination to discrimination}
So far, in the first experiment, we have looked at the locations of patterns yielded by different pattern extraction algorithms and explored some of the possible ways in which they might be combined.
To further investigate the musical events in these patterns, we move on to visualising their features and trying to use these to distinguish between them and human annotations.

\section{Visualising and classifying pattern features} \label{sec:class}
From our first experiment in combining the algorithms, we noticed that, to a large extent, the musical pattern discovery algorithms disagree between themselves as well as with human annotations.
However, the disagreement was down to the location of patterns rather than the actual content, namely the notes and rests in these passages.

In this section, we compare algorithms using the actual musical content rather than location of their output patterns.
By looking into the content of musical patterns, we can compare various musical features that might occur in the output of different algorithms.
We therefore extract features from the sequences of musical events that form these musical patterns and compare them between different algorithms and human annotations. 
Using dimensionality reduction techniques, we visualise these musical patterns in feature space, and perform Comparative Classification (\CC{}) between the algorithmically extracted patterns, human annotations, and random excerpts.
We then explain the classification results using feature importance analysis.

Note that analysing musical passages in isolation from their context might be limited in terms of generalisability, because a pattern in one musical piece might be insignificant in another piece depending on what surrounds it.
Nevertheless, certain arrangements of musical events and their features could be used to represent the way in which output differs between algorithms and from human annotations. 

\textbf{Classification based on content and performance on two subtasks} \label{yclass}
While it is evident from our results of our first experiment that the algorithms disagree on the locations of pattern occurrences, we did not consider and compare the actual musical content of the patterns. 
We might question, then, how the content of the patterns differs, especially in comparison to human annotations and to a random baseline. 

The comparison to the random baseline is of interest here because both human annotators and algorithms may produce output resembling randomness if, for example, they become confused by highly complex patterns in certain musical extracts.
To make such a comparison, we compute musical features from pattern occurrences and employ a range of classifiers to differentiate three groups of patterns: the algorithmically extracted patterns, the human-annotated patterns, and randomly sampled ``patterns'' (random excerpts).
To compare the individual algorithms, we also use classifiers to differentiate subgroups of patterns extracted by different algorithms. 
In other words, we perform \CC{} on two subtasks with different goals: the first
classification task is to attempt to classify patterns into three groups based on how they were extracted: algorithmically, manually, or randomly. 
In the second task, we perform a finer level of classification on the algorithmic group from the first task by attempting to classify patterns based on the algorithms that extracted them. 
In both tasks, we expect classifiers to help us discriminate between groups of patterns based on their musical features. 

\textbf{Data and algorithms}
For this second experiment, we use the \MTCANN{} Dutch Folk Song dataset~\cite{van2016meertens}, which has previously been used to compare pattern discovery algorithms.
The dataset contains an exceptionally large number of annotated patterns and is therefore suitable for use in our classification experiments.
During the making of \MTCANN{}, three experts were asked to annotate the prominent patterns in each song that align best with one of 26 tune families.
\textit{Tune family} is a concept in ethnomusicology that groups together tunes sharing the same ancestor in the process of oral transmission~\cite{cowdery1984tunef}.
The dataset consists of 360 Dutch folk songs with 1657 annotated pattern occurrences.
In an annotation study into which factors influence human judgement when categorising melodies belonging to the same tune family, repeated patterns were discovered to play the most important role~\cite{volk2012melodic}.
It is, therefore, reasonable to use repeated pattern discovery algorithms on this dataset. 
An example song from the dataset is shown in Figure~\ref{fig:examples}(b).

\begin{table}
 \begin{center}
 \begin{tabular}{l|c|c|c}
  \hline
  Algorithm & \#Pattern  & \#Occurrences  \\
  \hline
  SIACR & 893 & 5576 \\
  SIACP  & 250 & 3650 \\
  SIACF1  & 822 & 5308 \\
  VM  & 182 & 25679 \\
  VM2  & 159 & 4658 \\
  SC  & 126 & 355 \\
  SCFP  & 200 & 724 \\
  \hline
 \end{tabular}
\end{center}
 \caption{Algorithms and the number of patterns they extracted. 
 Explanations of abbreviations used are given in Section~\ref{subsec:visloc}. 
   We use the default configurations of these algorithms in this paper.
 }
 \label{tab:stats}
\end{table}

\textbf{Patterns from algorithms}
Table~\ref{tab:stats} shows the number of extracted patterns from state-of-the-art musical pattern discovery algorithms that have been used and compared in previous research~\cite{ren2017search,boot2016evaluating}.
We use the same set of algorithms in this paper. 

\textbf{Sampling random excerpts}
We compare human-annotated and algorithmically extracted patterns with randomly sampled excerpts as a baseline.
More specifically, the random excerpts are sampled using the following procedure:
\begin{enumerate}
    \item For each annotated pattern in \MTCANN{}, we find the corresponding song where the annotation appears.
    \item We then pick a random starting point and take an excerpt of the same length as the pattern to construct a candidate excerpt. 
    \item Finally, we repeat the sampling procedure five times to control for anomalous results.
\end{enumerate}

\textbf{Feature Calculation}
To obtain a feature representation of the musical pattern occurrences, we
calculate musically meaningful features using the prominent feature extraction tool, the jSymbolic2.2 toolbox in the j\MIR{} toolset~\cite{mckay2018jsymbolic}. 
jSymbolic2.2 takes a \MIDI{} file as input and from this file computes 246 unique musically meaningful features in eight categories: texture, rhythm, dynamics, pitch statistics, melodic intervals, chords and vertical intervals, instrumentation, and MEI-Specific features. 
Another well-known feature extraction package, the \FANTASTIC{} toolbox~\cite{mullensiefen2009fantastic} is not used because its minimum supported input length excludes valuable annotated patterns.
jSymbolic2.2 was not specifically designed for short excerpts and many human-annotated and algorithmically extracted patterns are indeed short in length. 
Nevertheless, jSymbolic2.2 computes features in the same way for all excerpts, it does not, therefore, undermine the validity of the classification experiments. 

\begin{figure}    
  \centering
  \captionsetup[subfloat]{farskip=0pt,captionskip=0pt}
  \vspace{-4mm}
  \subfloat[Feature importance for differentiating between annotation, algorithmic output, and random excerpts.]{{\includegraphics[width=0.8\columnwidth]{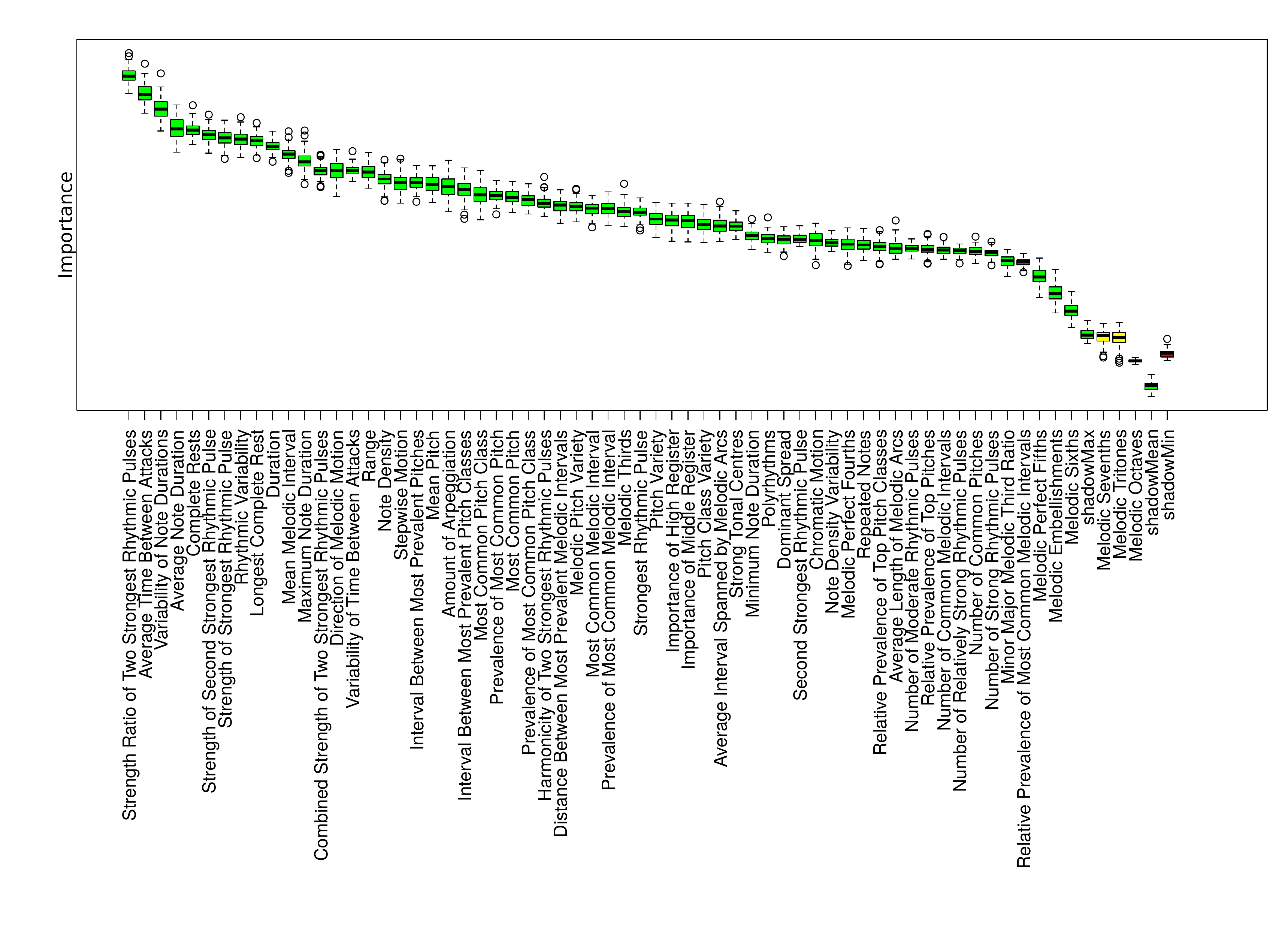} }}%
    \qquad
  \vspace{-5mm}
    \subfloat[Feature importance for differentiating between output from different algorithms.]{{\includegraphics[width=0.8\columnwidth]{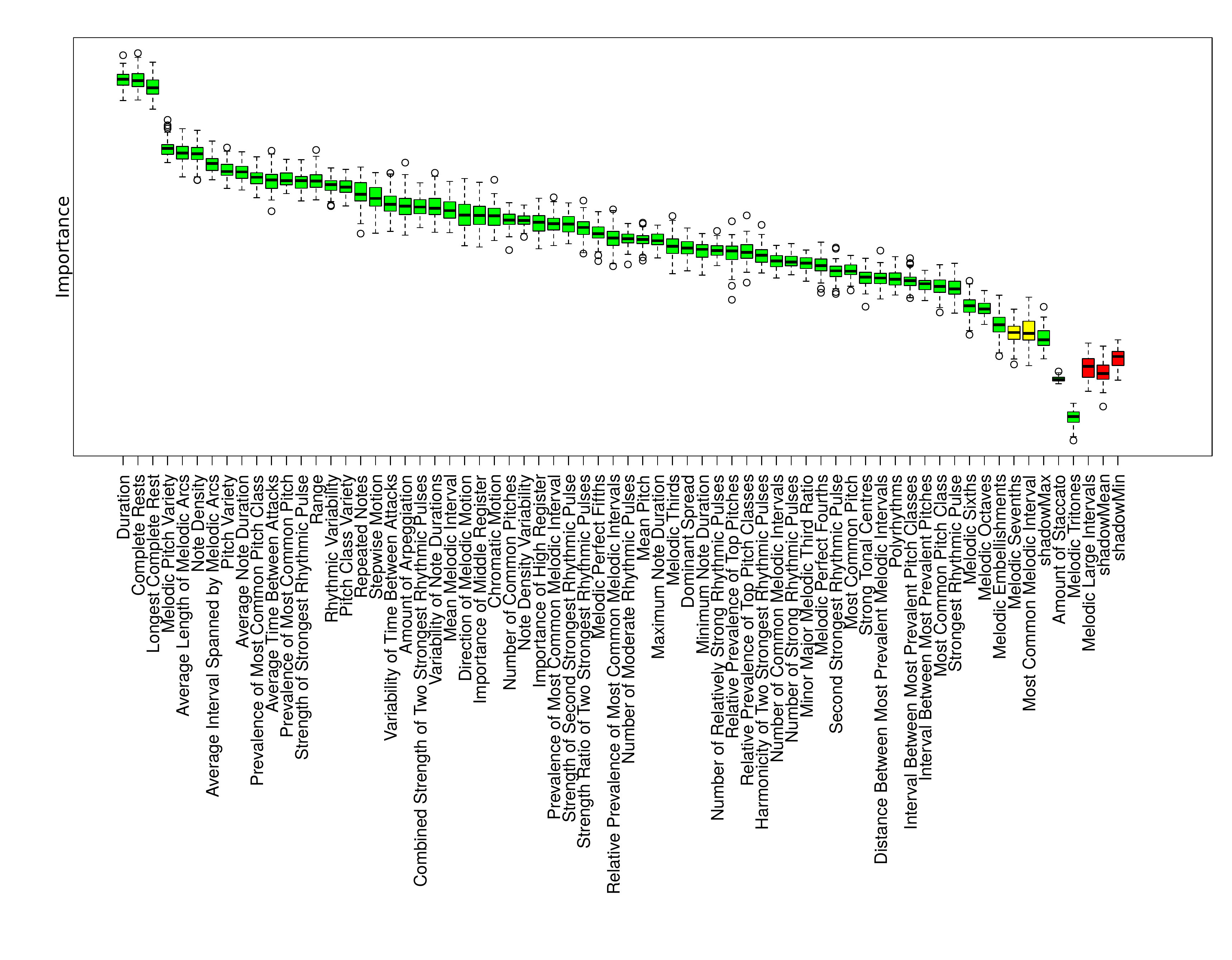} }}
  \vspace{+2mm}
 \caption{Feature importance in the two classification subtasks. 
 The boxplot shows the mean and variance (interquartile ranges) of the feature importance values~\cite{kursa2010feature}. 
 The features are ranked by their importance. 
 We omit the y-axis label because the absolute importance values are not relevant for our analysis. 
 The colour green indicates features that are more important than the randomised features and are therefore confirmed to be significant; blue entries show the performance of the random features; red and yellow indicate unimportant and tentative features respectively.}
 \label{fig:boru}
\end{figure}

\textbf{Feature Selection}
To avoid unnecessary computational effort and exclude irrelevant musical features, we perform a feature selection step. 
We first eliminate the features which are constant across all patterns, such as vibrato prevalence, average range of glissandos, etc. 
Next, we eliminate the features which are not relevant to time and pitch, such as the dynamics features and artefacts introduced by \MIDI{} conversion.
In the end, we retain 63 features as listed in Figure~\ref{fig:boru} and with which
we will have more detailed discussion at the end of this section.

\textbf{PCA and Visualisation}
We now visualise the three groups of patterns and their feature representations using Principal Component Analysis (\PCA{}).

\PCA{} is known to be a practical preprocessing step and dimension reduction strategy for visualisation and classification tasks. 
\PCA{} produces linear combinations of features which maximise variance in a given dataset and are therefore suitable for visualising differences within data.

\begin{figure}    
   \centerline{
 \includegraphics[width=0.9\textwidth]{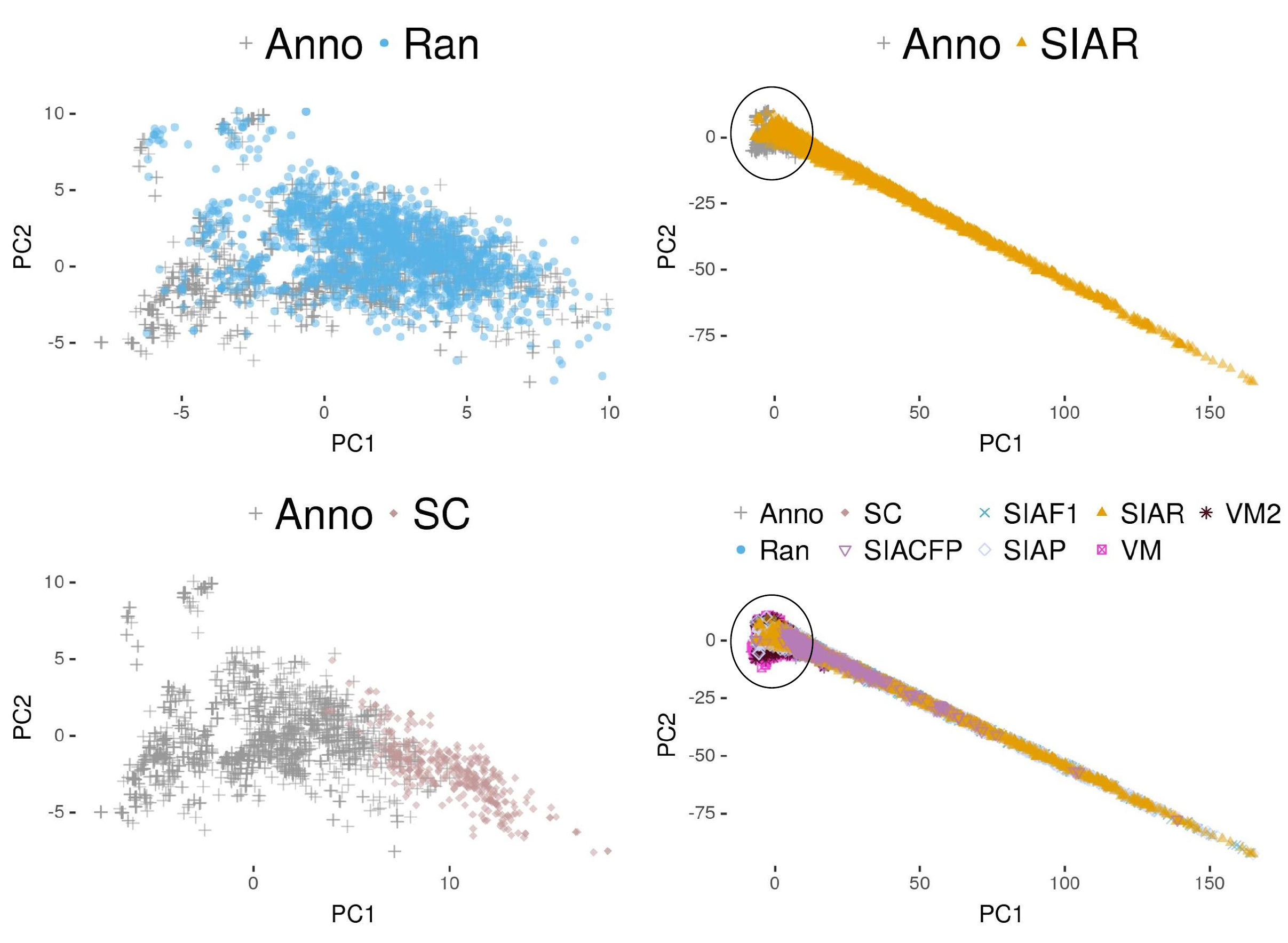}}
 \caption{Visualisation of different groups of patterns using the space spanned by the first two principal components of the annotated pattern features in \MTCANN{}. 
 The legend above each subfigure denotes the correspondence between colours and algorithmic patterns/annotated patterns/random excerpts. 
 Notice that the figures in the left column show zoomed subregions (circled) of figures in the right column. 
 (1) Upper left: random excerpts and annotated patterns. 
 The overlap between the two groups is large, and it is nontrivial to separate them in this two-dimensional \PCA{} embedding. 
 (2) Upper right: \SIACR{} patterns and the annotated patterns. \SIACR{} patterns exhibit a long tail which is not shared by the annotated patterns. 
 (3) Bottom left: \SC{} patterns and the annotated patterns. The overlap of the data points is small, which makes it easier to separate the two groups in this embedding. 
 (4) Bottom right: random excerpts, annotated patterns, and patterns from all algorithms. We can see that some of the algorithmically extracted patterns are very different from the annotated patterns, and the algorithms belonging to the same family exhibit similar long tails.}
 \label{fig:pca}
\end{figure}

In Figure~\ref{fig:pca}, we plot different groups and subgroups of patterns in a two-dimensional\footnote{More visualisations can be found at \texttt{https://goo.gl/qmyxdh}} \PCA{} embedding of the feature space. 
We make four cross-group comparisons to show typical cases of how musical patterns are distributed in the feature space spanned by the first two components of the \PCA{} decomposition. 
The visualisation is generated by using the annotated patterns as training data to obtain the \PCA{} embedding---the two principal components are used as the two axes of the feature space. 
Then, the random excerpts and patterns from different algorithms are projected onto this embedding.

From the four snapshots we take from the musical pattern \PCA{} feature space as shown in Figure~\ref{fig:pca}, we make several observations: 
\begin{itemize}
    \item  Annotated patterns and random excerpts have extensive areas of overlap, which makes it impossible to find a linear classifier that uses the first two principal components of the annotated pattern features, which in turn makes it nontrivial to differentiate the two groups of patterns as shown in the upper left subfigure. 
    \item \SIACR{} patterns exhibit a very different distribution from the annotated patterns and random excerpts shown in the top right subfigure. 
    Notice the annotated patterns concentrate at the top left corner. 
    In this case, it is relatively easy to separate the long-tail area of the extracted patterns from the annotation area. 
    By applying this observation and designing a filtering process, we could substantially improve the performance of the \SIACR{} algorithm on \MTCANN{}. 
    \item The overlap between the annotated patterns and extracted patterns is small in the bottom left subfigure. 
    A linear classifier can be devised to separate the two groups of data using the first two principal dimensions of the annotated patterns. 
    The extracted patterns of the \SC{} algorithm have different features to the annotated patterns. 
    \item In the bottom right subfigure, we show all the heterogeneous patterns as extracted by algorithms, annotated by humans, and randomly sampled in the same \PCA{} embedding. 
    Patterns extracted by algorithms of the same family, namely \SIACP{},  \SIACR{}, \SIACF{}, and \SIACFP{} tend to share the same long-tail property, and therefore their performance on \MTCANN{} can be improved by an extra filtering step as described above.
\end{itemize}
In summary, from a visual examination of Figure~\ref{fig:pca}, the application of classification techniques to discrimination between different groups of patterns based on their features seems like a promising approach to take.

\begin{figure}    
    \begin{center}
    \includegraphics[width=0.9\linewidth]{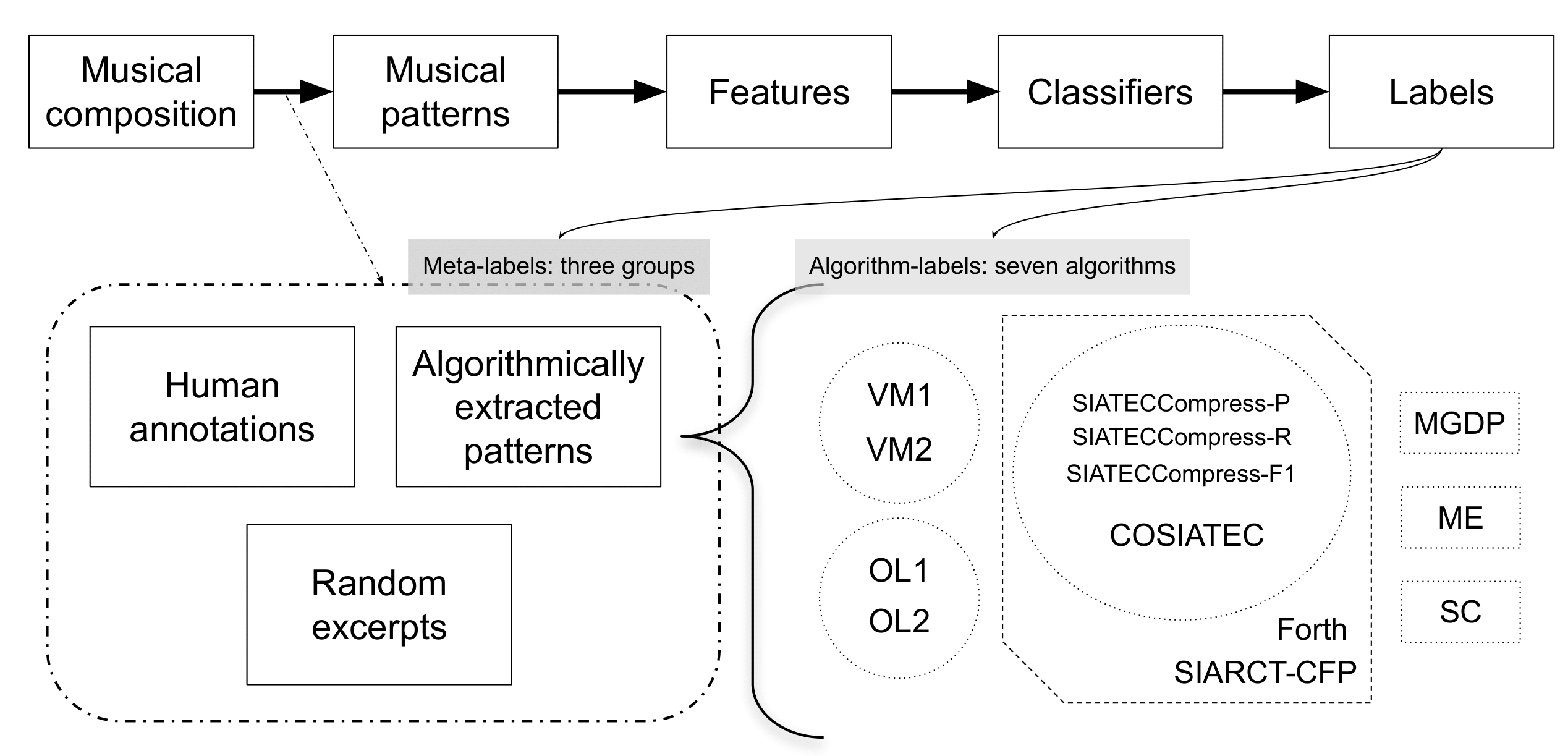}
  \end{center} 
  \caption{Pipeline of CC (the chain on top) and the two classification tasks
    using two types of labels. The meta-label classification experiment tries to
    use classifiers to distinguish between three groups of patterns from
    different origins. The algorithm-label classification experiment is
    performed on a finer level: we look into the algorithmically extracted
    patterns specifically and distinguish between seven subgroups of patterns.
    The seven algorithms we consider are in one of the oval shapes. The
    geometric methods are in the dashed line square with cut corners. Other
    algorithms are also listed to be considered in the next section. A short
    description of each algorithm can be found and matched in
    Section~\ref{sec:alg} and Table~\ref{tab:algs}. }
  \label{fig:classpipe}
\end{figure}

\textbf{Classification}
Supervised classification methods have been used extensively in \MIR{} tasks such as genre classification and the classification of corpora from different geographic origins. 
In addition to their use in labelling unknown data, classifiers can be used to examine the discriminating features between different groups of data.
In leveraging such discriminative power from classification algorithms, performing comparative analyses using classifiers is related to adversarial machine learning, and has been performed in other areas of research~\cite{da2009shape,patel2002analysis}. 
To the best of our knowledge, the pipeline we propose in Figure~\ref{fig:classpipe} has not yet been used to evaluate symbolic musical pattern discovery algorithms. 

In accordance with the two subtasks introduced earlier, we first use supervised classification methods to differentiate between human-annotated, algorithmically extracted, and randomly sampled excerpts in \MTCANN{}. 
Second, we use the same set of classifiers but with algorithmically extracted patterns from seven different algorithms to produce seven different subgroups of patterns.
By placing each pattern into one of three groups (the group of algorithmically extracted patterns, the group of human-annotated patterns, and the group of random excerpts) and each algorithmically extracted pattern into one of seven subgroups (extracted patterns from seven different algorithms), we observe whether or not the classifiers can detect systematic differences on the group and subgroup level. 

To prevent the results being classifier-specific, we use a mixture of simple and more sophisticated, linear and non-linear classifiers to perform the classification tasks. 
We also use standard machine learning techniques to train and test classifiers: scaling and centring preprocessing steps are first performed on all the features and \PCA{} input; additionally, to avoid overfitting, for all experiments, we use a 10-fold cross-validation 3-times repetition scheme. 
The \PCA{} projection and parameter search for each classifier are performed separately on each fold. 
The six statistical classifiers we use are:

\textbf{GBM}~\cite{friedman2001greedy} (Gradient Boosting Machine) produces a prediction model consisting of an ensemble of decision trees. 
The parameters we search through are the learning rate, complexity of trees, minimum number of samples to commence splitting, and the number of iterations.

\textbf{LVQ}~\cite{kohonen1990improved} (Linear Vector Quantisation) applies a winner-takes-all Hebbian learning-based approach. 
We search through two parameters in this classifier: codebook size and number of prototypes.

\textbf{LDA}~\cite{ripley2007pattern} (Linear Discriminant Analysis) produces a linear classifier which finds a linear combination of features that best separates different classes in datasets. 
This classifier is not parametric.

\textbf{NB}~\cite{ng2002discriminative} (Naive Bayes) computes the conditional a-posterior probabilities of a categorical class variable given independent predictor variables using the Bayes rule. 
Three parameters are tuned for this classifier: Laplace smoothing, kernel bandwidth, and distribution type.

\textbf{RF}~\cite{breiman2001random} (Random Forest) operates by constructing a multitude of decision trees. 
The parameter we consider is the number of variables per level.

\textbf{SVM}~\cite{suykens1999least} (Support Vector Machine) calculates a map from data to a new representation so that the data points of the separate categories are divided by a gap that is as wide as possible. 
We use the radial basis function kernel and consider two parameters: smoothing factor and weight of training examples.

\textbf{Results}
We mainly use accuracy and the variance of accuracy as the measure of the performance of the classifiers. 
To further interpret the results of the classification task, we compute confusion matrices and feature importance measures. 
Ten other metrics for each classifier are provided for further inspection\footnote{\texttt{https://goo.gl/ezuTCT}}. 
Below, we only report the most relevant results.

\begin{figure}    
  \centering
  \subfloat[Accuracy for the three groups classification]{{\includegraphics[width=0.4\columnwidth]{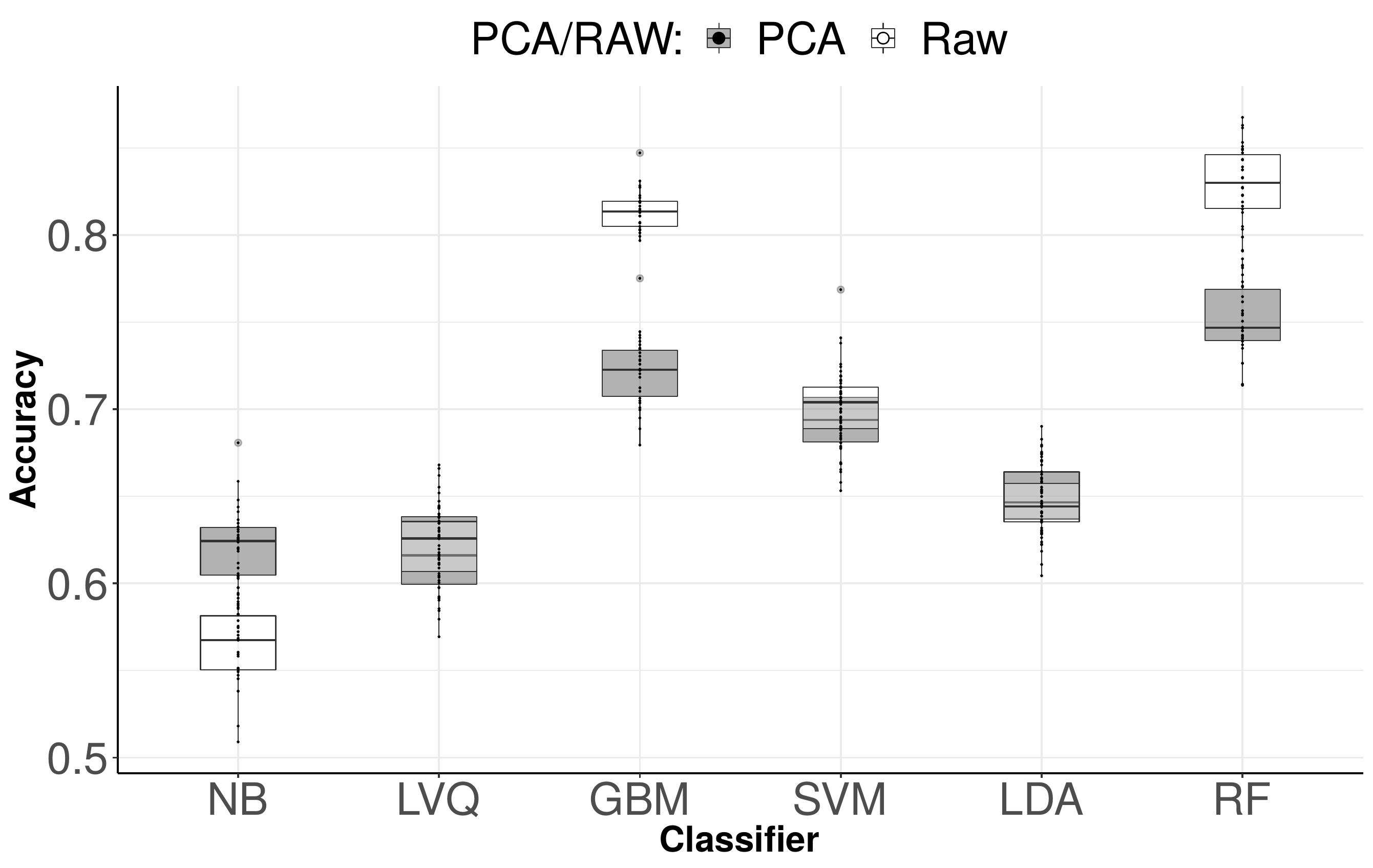} }}
    \qquad
  \subfloat[Accuracy for the seven algorithms classification]{{\includegraphics[width=0.4\columnwidth]{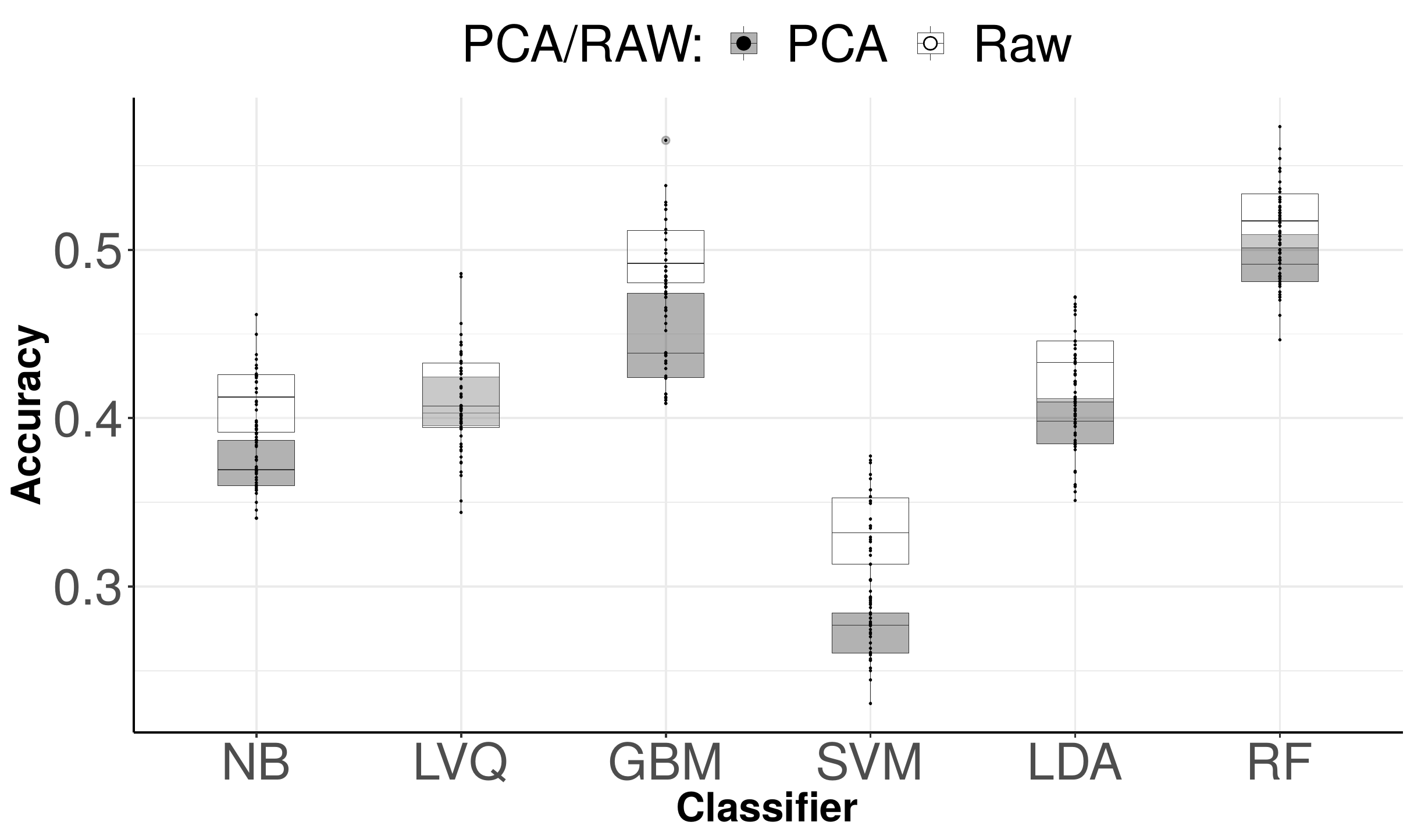} }}
 \caption{Accuracy values for classifiers in thirty experiments (10-fold cross-validation repeated three times) using six classifiers with j\textsc{s}ymbolic\oldstylenums{2}.\oldstylenums{2} features and after \textsc{pca} preprocessing.}
 \label{fig:accu}
\end{figure}

\textbf{Model metrics}
In Figure~\ref{fig:accu}, we show the accuracy and variance of different classifiers in the two classification tasks. 
We use two groups of features, the raw features and features after the \PCA{} preprocessing. 
The baseline accuracy is $\frac{1}{\#group} = \frac{1}{3} = \sim 33\%$ for the first task and $\frac{1}{\#group} = \frac{1}{7} = \sim 14\%$ for the second. 
We balanced the number of patterns in each group to make them the same, for the first task $=1657$, and for the second task $=355$.

We see that all classifiers give a result higher than the baseline accuracy in both tasks. 
\PCA{} improves the performance of the classifier \NB{} in the first task.
For \LVQ{}, \SVM{}, and \LDA{}, using \PCA{} or raw input does not make a significant difference.
The performance of other classifiers is worse when using the \PCA{} input. 
The finding that \PCA{} has different influences on the performance of classifiers may be due to the fact that each classifier uses different internal feature transformation mechanisms. 
Overall, the random forest classifier gives the best results for both tasks with the raw feature input and the parameter $\#variables=32$.

The fact that we can construct a classifier to differentiate between groups of patterns with high accuracy has a few implications. 
In the first task, it implies that algorithmically extracted patterns possess different properties than human-annotated patterns, which suggests that extra consideration of features of patterns when trying to discover patterns automatically would be beneficial; 
it also shows that the algorithmically extracted patterns have different traits than random excerpts, which means that these patterns are not equivalent to randomly sampled excerpts, and are therefore potentially more useful for some applications; 
lastly, it shows a difference between human-annotated patterns and randomness despite subjectivity being involved in the annotation process, which is in agreement with the carefully designed annotation collection process~\cite{van2016meertens} and the previous findings that the annotations are useful for classifying tune families~\cite{boot2016evaluating}. 
In the second task, the above-baseline accuracy shows distinguishability in the patterns extracted by the algorithms, which suggests disagreement between different algorithms, reinforcing our conclusions from the first experiment.
To further investigate differences between groups, we now analyse the confusion matrices. 

\textbf{Confusion Matrices}
In Table \ref{tab:cm3} and \ref{tab:cm7}, we give the confusion matrix results calculated from the classifier that has the best classification results: Random Forest. 
We perform the repeated cross-validation experiment ten times and take the mean and variance of the resulting ten confusion matrices. 
The results show us that different groups of patterns are separable and dissimilar to one another according to the random forest classifier. 

To read the table, we first notice that the sum of each column is roughly 500, which is the size of our test data. 
The row sums do not have this constraint because we do not place restriction on the group size as determined by the classifier. 
For interpreting the entries in the table, we take as an example the number 29.5 in the top right corner of the table.
This number is the mean count of patterns classified as algorithmically extracted but are actually annotations. 
Table~\ref{tab:cm7} is formatted similarly, with a different column sum because of the different test data size, and with algorithm names instead of the group names. 

\begin{table}
\centering
\begin{tabular}{cccc}
  \hline
 \makecell{Original $\rightarrow$ \\ Classified $\downarrow$}  & Alg & Ran & Anno \\
  \hline
  Alg & 406 ($\pm$13.5)  & 32.3 ($\pm$7.2)  & 29.5 ($\pm$8.3)  \\ 
  Ran & 54.1 ($\pm$8.9)  & 402 ($\pm$15.9)  & 65 ($\pm$14.1)  \\ 
  Anno & 37 ($\pm$8.4)  & 62.8 ($\pm$11.3)  & 402.6 ($\pm$14.2)  \\ 
  
   \hline
\end{tabular}
\caption{Confusion matrix results from the ternary classification experiment using the Random Forest classifier: mean and variance (in parenthesis) of ten experiments. 
The row names indicate that the patterns are classified into the group with this name by the classifier; the column names indicate the patterns are originally from the group with this name. 
Three groups of data are classified with high accuracy and significant p-values $\ll$ 0.05. }
 \label{tab:cm3}
\end{table}

\begin{table}
  \small
\centering
\begin{tabular}{rlllllll}
  \hline
\makecell{Original $\rightarrow$ \\ Classified $\downarrow$} & SIACF1 & SIACP & SIACR & VM2 & VM1 & SC & CFP \\ 
  \hline
SIACF1 & 22.6($\pm$9.7)  & 18.6($\pm$9.9)  & 23.6($\pm$12.8)  & 2.5($\pm$2.3)  & 0.4($\pm$0.8)  & 1.4($\pm$1.5)  & 8.4($\pm$4.7)  \\ 
  SIACP & 12.8($\pm$6.8)  & 34.4($\pm$6.6)  & 15($\pm$5.9)  & 0.5($\pm$0.9)  & 1.3($\pm$1.8)  & 1($\pm$1.9)  & 5.1($\pm$2.5)  \\ 
  SIACR & 30.6($\pm$12.1)  & 23($\pm$8.7)  & 31.8($\pm$9)  & 1.8($\pm$2.1)  & 1($\pm$1.3)  & 1.3($\pm$1.9)  & 10.6($\pm$5.6)  \\ 
  VM2 & 8.2($\pm$5.1)  & 11.1($\pm$4.2)  & 9($\pm$4.4)  & 63.1($\pm$7.2)  & 21.6($\pm$7.8)  & 0.3($\pm$0.6)  & 3.8($\pm$2.5)  \\ 
  VM1 & 0.9($\pm$1.4)  & 5.5($\pm$2.9)  & 1.9($\pm$2.2)  & 22.5($\pm$6.9)  & 78.5($\pm$8.8)  & 0($\pm$0)  & 0.3($\pm$0.9)  \\ 
  SC & 16.2($\pm$6.5)  & 6.8($\pm$4.2)  & 10.3($\pm$4.8)  & 7.3($\pm$3.9)  & 0.2($\pm$0.6)  & 93.8($\pm$4.9)  & 17.1($\pm$4.1)  \\ 
  CFP & 15.2($\pm$5.9)  & 7.2($\pm$4)  & 14.9($\pm$5.6)  & 8.7($\pm$4.6)  & 3.6($\pm$3.1)  & 8.8($\pm$4.5)  & 61.1($\pm$9.2)  \\
   \hline
\end{tabular}
\caption{Confusion matrix results classifying the algorithmic output using the Random Forest classifier. 
The columns and rows follow the same format as in Table~\ref{tab:cm3}. }
\label{tab:cm7}
\end{table}

In the first task, we see that the classifier can differentiate between the three groups with few instances of incorrect classification.
This is not the most desirable result for the pattern discovery algorithms. 
If we had seen that the classifiers could not differentiate between the algorithmically extracted pattern group and the human-annotated pattern group, this would suggest a level of consensus between algorithms and humans. 
In other words, if the count is larger at the last line of the first column in the confusion matrix, it would indicate that the patterns discovered by algorithms are harder to distinguish from the human annotations, which is more desirable if the algorithms would like to imitate the pattern discovery behaviours of human annotators. 
With the current results, however, we come to the conclusion that the algorithmically extracted patterns, annotated patterns, and random excerpts possess their own traits and are not similar enough for the classifiers to fail. 
This is in accordance with the first experiment and previous research that the extracted patterns are not yet indistinguishable from the human annotations~\cite{boot2016evaluating, ren2017search}. 
Despite this, we at least establish that neither annotated patterns nor extracted patterns are equivalent to random data.

In the second task, the classifier can also differentiate patterns from different algorithms to a certain extent. 
The goal in this task is to examine the individual pattern discovery algorithms.
Similarly to the analysis in the first subtask, if we can distinguish between different algorithms perfectly, this would indicate that the discovered patterns are very different, despite the algorithms having the same objective of ``pattern discovery''; and if the algorithms are in agreement with each other, we would expect the classifier to fail and the values in the confusion matrix to be more uniformly distributed. 
We see in Table~\ref{tab:cm7}, however, this is often not the case.
The exceptions here are the patterns extracted by algorithms from the same family (\SIACF{}, \SIACP{}, \SIACR{} for example): the classifier performs worse in distinguishing within the same family. 
This is an indication that we are not overfitting the classifier.
More importantly, this classification result agrees with the first data fusion experiment in that the output from algorithms has a high degree of disagreement, especially when the algorithms come from different families. 

\textbf{Feature Importance}
In Figure~\ref{fig:boru}, we show the individual importance value of each feature used in the random forest algorithm (the best classifier in the two classification subtasks), by using the Boruta algorithm~\cite{kursa2010feature}. 
The Boruta algorithm randomly duplicates and shuffles the values in the original features as extracted by jSymbolic. 
The algorithm then combines some of these randomised features with the originals in order to calculate and compare the significance of different combinations of features. 
At the end of this process, we obtain an importance value for each feature; specifically the Gini impurity importance value~\cite{louppe2013understanding}.


Although there are only 23 rhythmic features out of 63 in total, in both classification subtasks, the features ranked highest in significance were rhythmic in nature.
This suggests that these features were more important than other features in constructing the random forest classifiers, which hints at potential improvements that could be implemented for current existing pattern discovery algorithms. 
String-based and data mining algorithms translate pitch and duration pairs into a list of symbols and therefore do not take into account metric structures imposed by musical punctuation such as bar lines. 
This is not uncommon as other algorithms also seldom explicitly consider metric features in patterns. 
In particular, the \SC{} algorithm we use considers pitch aspects only, which provides an explanation for its limited overlap with human-annotated patterns in the \PCA{} visualisation.  
Generally, the feature importance values we obtained suggest that in designing and evaluating pattern discovery algorithms, at least for the \MTCANN{} dataset, we should take metric structures into consideration as well as repetition and pitch related features in patterns.

Our findings suggest that other jSymbolic features are also important, with the exception of three, which performed worse or at the same level as randomised features. 
The Boruta algorithm categorises Melodic Octaves feature as unimportant and the Melodic Sevenths and Melodic Tritones as tentative features. 
They are nonessential features because these musical intervals rarely appear in the \MTCANN{} dataset.

\textbf{Discussion}
In the second experiement, we visualised and distinguished between human-annotated patterns, algorithmically extracted patterns, and random excerpts in \MTCANN{}.
In performing \CC{} in two subtasks, we obtain results suggesting that algorithmically extracted patterns can still be distinguished from human-annotated patterns, and that both algorithmically extracted and human-annotated patterns show different traits than randomly sampled excerpts.
The extracted patterns from different algorithms can also be distinguished, partly due to the manner in which they disagree with each other. 
\textsc{cc} provides a new means of examining the output of pattern discovery algorithms, in addition to comparison of individual patterns in the \MIREX{} task and in the first experiment in Section~\ref{sec:ismir}.
A subsequent feature importance analysis reveals that rhythmic features play an important role in distinguishing between the groups of patterns in \MTCANN{}.

\section{Using synthetic data for evaluation}\label{sec:synth}
\textbf{Previous two experiments}
In Section~\ref{sec:ismir}, we demonstrate that algorithms often disagree with humans, as well as between themselves, on what constitutes a pattern, whether this is in the \JKUPDD{} or the \MTCANN{} dataset.
To understand where and how the algorithmic output disagrees, we computationally examined the musical patterns by visualising, combining, and classifying them.
Through this algorithm-algorithm and algorithm-human comparison, we see that a diverse range of patterns are extracted from algorithms and annotated by humans.

\textbf{Building on our findings}
We can see from Figure~\ref{fig:examplehanon} and Figure~\ref{fig:examplemozart} that it is possible to perceive and discuss patterns at different levels of musical structure, which can themselves be contingent on a listener's general experience with music. 
Such diversity of musical patterns creates issues in \textit{defining} what musical patterns are and \textit{select} the algorithms with respect to the extent to which their output reflects human annotations, contributing to the difficulty of producing algorithms that yield high-quality output.
By taking a high-level overview of the disagreements between algorithms and the factors that contribute to them, it is easy to motivate the search for methods to inspect, compare, validate, and select these algorithms.
More specifically, we are motivated to search for a representation sufficient to capture the characteristics of a large number of extracted patterns (inspect), to quantify the differences in patterns extracted by different algorithms (compare), to determine that the output matches pattern extraction goals (validate), and choose the most suitable algorithms for the use case (select).
While \PP{}, \CC{}, and the visualisation methods in the previous two experiments constitute our first attempt to realise some of these steps, we still lack an intuitive understanding of how these algorithms could perform given new musical corpora. 

\textbf{Approach}
Given the complexity of some musical corpora, we propose the use of synthetic data in conducting an initial assessment of the performance of musical pattern discovery algorithms. 
For example, one of the simplest and yet most interesting questions to ask is whether pattern discovery algorithms are capable of identifying patterns in sequences such as ``Pattern1'' + ``Pattern2'' + ``Random Excerpts'' + ``Pattern1''. 
By artificially constructing a concatenation of musical patterns and random sequences of notes and rests, we can compare the performance of different algorithms at extracting patterns from sequences such as this.
That is to say, random sequences with patterns artificially inserted throughout, giving a controlled amount of regularity. 
This is a method which has been widely used in other areas such as generic pattern mining and time series analysis~\cite{bertens2016keeping, Chiu2003tsmotifs}.
To the best of our knowledge, it has not yet been used extensively to compare musical pattern discovery algorithms.

\textbf{Evaluating algorithms}
In this section, we examine seven musical pattern discovery algorithms using
synthetic data: \SIATECC{}ompress - \SIACP{} (\SIACP{}), \SIATECC{}ompress -
\SIACF (\SIACF), \SIATECC{}ompress - \SIACR{}(\SIACR{})~\cite{meredith2016using}, \VMM{} \& \VMMM{}~\cite{velarde2016wavelet}, \MGDP{}~\cite{conklin2010discovery}, and Forth~\cite{forth2012cognitively}.
Because not all algorithms are open sourced, we are only able to obtain the patterns extracted by algorithms on certain datasets. 
Please refer to Table~\ref{tab:algs} for the algorithms we added and removed from previous experiments in this section.
In the following, we first describe how we created the synthetic data and then show an example piece from which different algorithms extract different patterns. 

\textbf{Planted patterns and randomness}
To create the synthetic data, we randomly concatenated two predefined patterns with random excerpts.
Given that we have planted the patterns artificially, we are able to compare the output of the algorithms to what is ostensibly a ``ground truth''.
The details of the predefined patterns are given next. 

\begin{figure}    
  \makebox[\textwidth][c]{\includegraphics[width=1\textwidth]{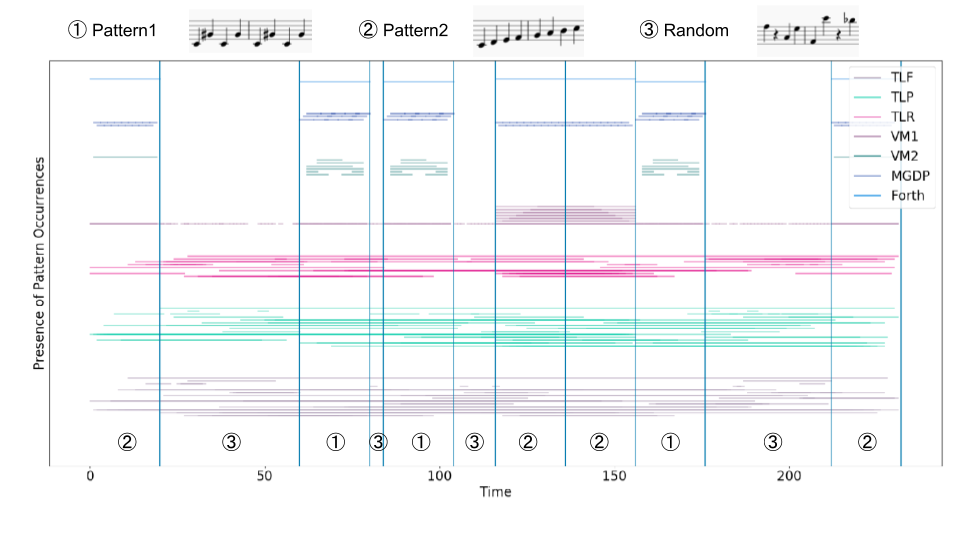}}%
  \caption{Visualisation of the presence of patterns using synthetic data by all algorithms. }
  \label{fig:simu}
\end{figure}

\textbf{Two patterns}
We use two predefined patterns in our synthetic data.
The excepts of the patterns can be found in Figure~\ref{fig:simu}.
In more detail, one pattern is a consecutive repetition of the notes C and \textsharp{G} in the fourth octave.
We refer to this pattern as $P_1$, the repeated interval pattern.
The other pattern we use is a C major scale, also known as the C Ionian scale in a modal context.
We refer to this pattern as $P_2$, the scale pattern.
Both patterns are twenty notes long.
The scale pattern will go on for three more bars in Figure~\ref{fig:simu}, and the same applies to the interval pattern. 
The reason we chose these two patterns is because they are of different kinds: $P_1$ contains many local repetitions, while $P_2$ is an example of a global pattern with a longer span. 

\textbf{Random excerpts}
We sample the random excerpts, $R$, with the same note range as the scale pattern.
We sample rests as well as notes. 
The lengths of the random excerpt are not fixed, but we constrain the total length of the random excerpts to be less than 50\% of the total length of the synthetic piece.

\textbf{Rationale}
Because we would like to put more focus on investigating how pattern discovery algorithms detect pitch related patterns, every note and rest in $R$, $P_1$, and $P_2$ has the note length of a crotchet.
Repeated interval patterns such as $P_1$ can be used to test whether the algorithms can retrieve local repetitions of intervals, and scale patterns such as $P_2$ can be used to test whether the algorithms can identify global repetitions of an organised sequence of notes.

\begin{figure}    
    \makebox[\textwidth][c]{\includegraphics[width=1\textwidth]{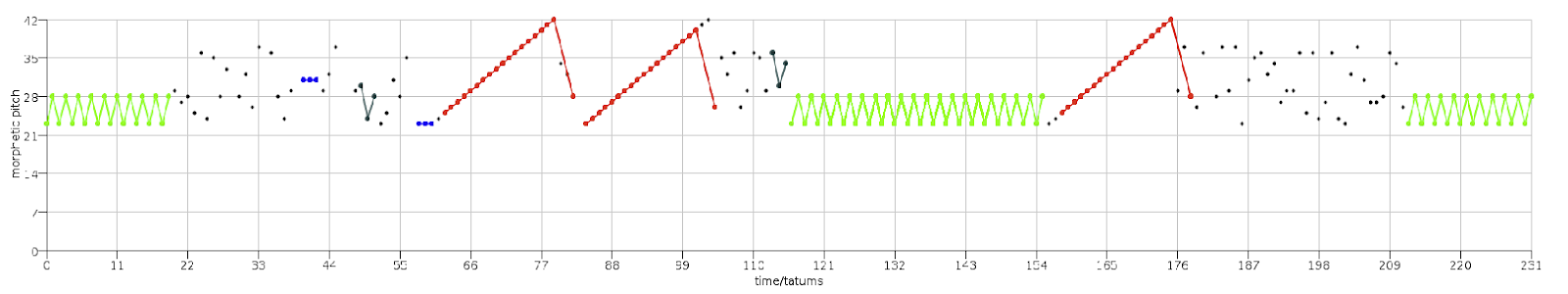}}%
  \caption{Patterns retrieved by the \textsc{siarct} algorithm. The x-axis represents time and the y-axis pitch. Each point represents the onset of a note. Different colours highlight different pattern occurrences.  The same applies to Figure~\ref{fig:forth} and Figure~\ref{fig:cosia}}
 \label{fig:siarct}
\end{figure}

\begin{figure}    
    \makebox[\textwidth][c]{\includegraphics[width=1\textwidth]{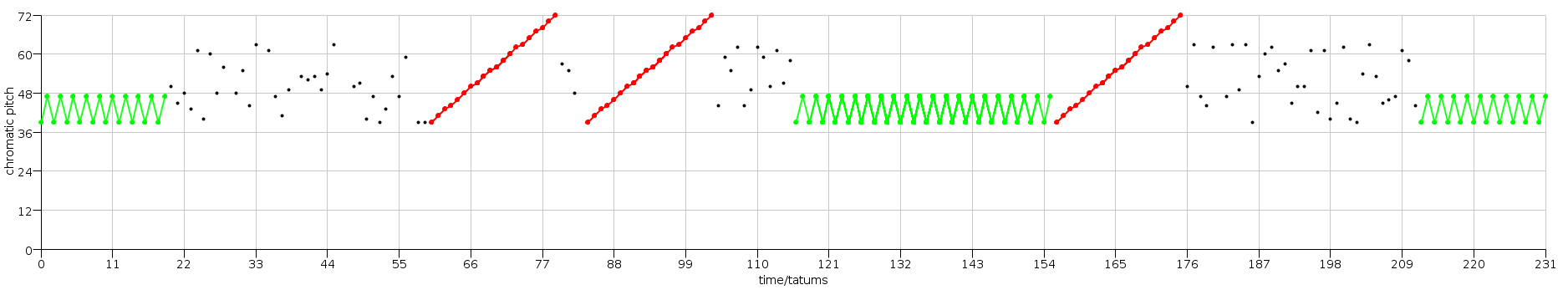}}%
  \caption{Patterns retrieved by the Forth algorithm. }
  \label{fig:forth}
\end{figure}

\begin{figure}    
    \makebox[\textwidth][c]{\includegraphics[width=1\textwidth]{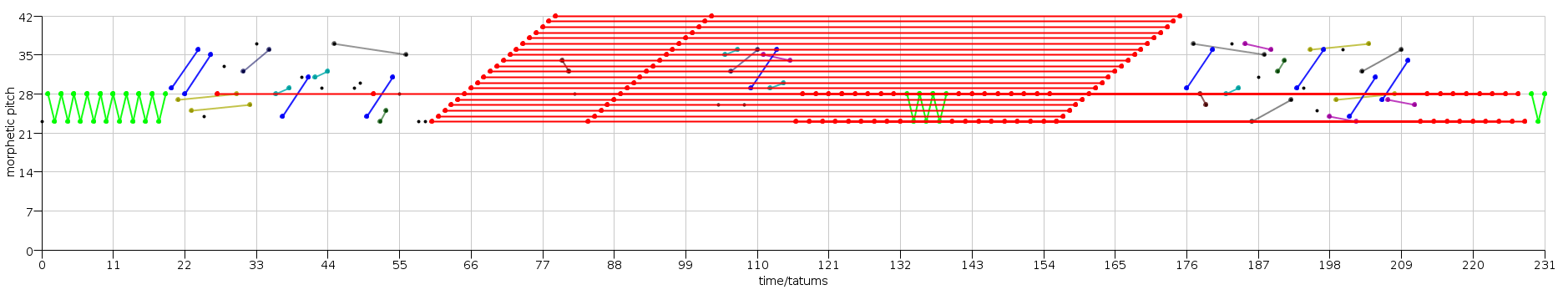}}%
  \caption{Patterns retrieved by the \COSIATEC{} algorithm. }
  \label{fig:cosia}
\end{figure}

\textbf{An example}
We present one piece from our set of synthesised pieces to illustrate the differences in the output produced by the algorithms. 
This piece can be seen in Figure~\ref{fig:siarct},~\ref{fig:forth}, and~\ref{fig:cosia}. 
We further discuss the pattern discovery results in the coming paragraphs. 

\textbf{Results}
Figure~\ref{fig:simu} visualises the presence of discovered patterns along the time axis of the synthetic piece, as introduced in Section~\ref{sec:ismir}.
We discuss our observations with respect to individual algorithms below. 

\textbf{VM}
The ground truth and the discovered patterns are colour coded.
We observe that \VMM{} consistently finds $P_2$, and \VMMM{} likewise for $P_1$.
However, the algorithms discover multiple subpatterns in the regions where $P_1$ and $P_2$ are present.
Regarding the \PP{} method introduced in Section~\ref{sec:ismir}, if we summarise algorithmic output by summing the counts of the discovered patterns, the resulting polling curve will give a correct indication as to their presence.
This tells us that the algorithms could benefit from a combined approach, though only in specific circumstances. 

\textbf{MGDP}
Patterns discovered by the \MGDP{} algorithm are shorter patterns.
The parameters used were $\alpha=0.1$, $Min(support)=20$, $Viewpoint=Interval$.
The discovered interval patterns are $\{8,-8,8\}$, $\{-8,8,-8\}$, $\{2,2,1\}$, $\{1,2,2\}$, $\{2,1,2\}$.
We can see from Figure~\ref{fig:simu} that the trigram (three-component) interval patterns cover $P_1$ and $P_2$ completely, though with multiple overlapping occurrences.

\textbf{Forth}
The Forth algorithm successfully retrieved all the planted patterns in the example piece.
As we can see in Figure~\ref{fig:simu}, by comparison with the ground truth (numbered \textcircled{1}-\textcircled{3}), the two planted patterns and their occurrences are all retrieved.
In Figure~\ref{fig:forth}, we can see in more detail that the algorithms find both the repeated intervals and the scales. 

\textbf{SIARCT}
As shown in Figure~\ref{fig:siarct}, while patterns retrieved by \SIARCT{} are indeed repeated patterns in the piece, they tend to deviate from $P_1$ and $P_2$. 
More specially, the scale patterns are corrupted by adjacent notes in the random sequences.
In addition, the algorithm finds a small number of patterns in the random pitch sequences.
Nevertheless, the approximate structure of the piece is retrieved. 

\textbf{SIACPRF}
Without parameter optimisation and applying out-of-the-box \SIACR{}, \SIACP{}, and \SIACF{} algorithms, we find many patterns in addition to the ones that are injected into the piece in Figure~\ref{fig:simu}.
The output does not contain occurrences that are confined to within the boundaries of the planted patterns, as was the case for \VMM{} and \VMMM{}.
Not only this, but most of the extracted patterns cover a span of time in the piece that is much longer than the planted patterns. 

\textbf{Patterns with long spans}
Figure~\ref{fig:cosia} shows long-span patterns which, with a concrete view of the example piece, can be seen to be unintuitive but not unreasonable: the injected scale $P_2$ repeats across the piece, but the elements within the scales also repeat individually.
This repetition is less intuitive because of limited human memory and the Gestalt principle of proximity in grouping notes~\cite{lerdahl1985generative,snyder2000music}---it is more plausible for an audience to identify the scale as a musical pattern rather than each individual note over a long time span.
There is a duality between the locally bounded repetition (where the scale repeats as a whole) and globally identifiable repetition (notes in scale repeat separately).
Depending on the subsequent applications of the extracted patterns, the algorithm should give more consideration to balancing or filtering out patterns of the desired kind.
For example, if the patterns are to be used to assist with musical motif analysis, the local patterns are more valuable, if our goal is to uncover hidden structures in music, global patterns could open up more possibilities and insights.

\textbf{Comparison with ground truth}
As we have discussed regarding long-span patterns, it would be hard to automatically evaluate the patterns discovered by \SIACR{}, \SIACP{}, and \SIACF{}, because there can be multiple patterns extracted from the constructed piece, and it is not fair to determine the correctness of the discovered patterns without application scenarios in mind. 
In addition, we must also consider the issue of intentionality in the use and recognition of planted patterns. 
On one hand, an unintended consequence of the concatenation process might be the emergence of new patterns. 
It is therefore possible to inadvertently introduce inexact and exact repetitive patterns in addition to those we planted deliberately.
On the other hand, the two patterns we employed are distinctive enough that it is likely that a human annotator would be able to find the exact boundaries of both.
Nevertheless, there are known cognitive phenomena such as apophenia~\cite{steyerl2016sea}, patternicity~\cite{shermer2008patternicity}, and hyperactive agency detection~\cite{valdesolo2014awe}.
If there is a tendency to find meaningful patterns in meaningless noise~\cite{fyfe2008apophenia, shermer2008patternicity} in human subjective experience, the question of how one should evaluate the algorithmic output in relation to the ground truth human annotations remains a complex one.
Therefore, despite the differences we observe in all three experiments, we cannot repudiate the value of the discovered patterns from algorithms.

\textbf{Rhythmic features eliminated}
Notice that, in the example above, we reduced the variance in rhythmic features by using the same duration for each musical event.
Rhythmic features, being the most divergence-inducing factor in \CC{}, should give the algorithms more tendency to conform, once we control their variance by artificially increasing their regularity.
Adding in more rhythmic variation could lead to greater insight as to how the algorithms handle the interplay between pitch and rhythm. 
We defer a more systematic investigation on this topic to later work.
The minimal example we provided, despite being restricted to pitch, informs us on crucial aspects of the algorithms with respect to the pitch dimension. 
This is another advantage of using synthetic data, which gives us a higher level of control in investigating musical features. 

\textbf{Implications for the human-algorithm gap in musical pattern discovery}
By examining the patterns extracted from the synthetic data in this section, we gain new insight into how we expect the algorithms to perform given controlled input data.
For example, if we apply pattern discovery algorithms to a musical piece with sections of ostinatos ($P_1$ being a special case) and ornamental bridge ($P_2$ being a special case), the algorithms might not be able to return the units in the two sections as patterns.
This might come as unexpected to human annotators who are looking for obvious repetitions in the piece. 
By reducing the complexity of current patterns discovery corpora to simple concatenations of preselected patterns, synthetic data is potentially helpful for better inspecting, comparing, validating, and evaluating the algorithms by modelling the most likely behaviours of human annotators.  
For example, one can identify which algorithm is better suited than others in retrieving ``conventional'' (exact repetition) or ``surprising'' (regularities in random) kind of patterns.

\section{Concluding remarks} \label{sec:dis}
\textbf{Summary: challenges and contributions}
In this paper, we devised new methods and evaluated musical pattern discovery algorithms computationally in three new ways: pattern polling (\PP{}), comparative classification (\CC{}), and synthetic data with planted patterns.
These new methods serve the purpose to better enable the inspection, comparison, validation, and evaluation of the algorithms beyond the limitations imposed by using numerical evaluation metrics such as the accuracy and the F1-score. 
In addition to extending previous evaluation methods on pattern algorithms, we discussed the challenges posed by the ambiguity and complexity in the musical corpora, which require different approaches to unravelling these different interpretations and characteristics of patterns in these corpora.
Another layer of uncertainty for designing pattern discovery algorithms comes from the diversity and subjectivity in the perception of patterns, which makes an all-encompassing definition of musical patterns too vague, and a highly specific one too narrow and inaccessible.
All these factors complicate the design of the methods for efficiently inspecting, intuitively comparing, constructively validating, and effectively selecting the algorithms. 
\PP{}, \CC{}, synthetic data with planted patterns are approaches that could address these problems.

More specifically, as shown in the \PP{} section, we can visualise and inspect the large number of algorithmically discovered or annotated patterns based on the locations of their occurrences. 
Based on this visualisation, the polling curve shows the fluctuations in how many patterns are discovered by the algorithms at a given time offset in music. 
From calculating the critical points on the polling curve, boundaries of pattern occurrences can be obtained when there is agreement between algorithms. 

Alternatively or additionally, we can compare the pattern occurrences using their musical content.
In \CC{}, by considering a range of features from different groups of patterns, we compute and represent each pattern occurrence in a feature space.
The subsequent visual inspection and classification expose dissimilarities in different types of features and therefore have the potential to provide high-level directions for improving algorithms. 
For example, in \MTCANN{}, rhythmic aspects have the most influence on how well the classifiers perform at distinguishing between different groups of patterns, which suggests a stronger consideration (of) the usage of rhythmic aspects in musical pattern discovery. 

Based on these findings of the varied and complex output from algorithms, we then proposed the use of synthetic data with planted patterns, evaluating the pattern discovery algorithms in a simpler and more controlled manner. 
By giving an instance of the synthetic data, we found that some less obvious patterns are extracted by the algorithms as well as the patterns we planted. 
This provides grounds for users of the algorithms to validate their hypotheses about the algorithms and make choices based on their particular use cases.

Evaluation results of \PP{}, \CC{}, and synthetic data vary across the limited datasets available, and these results might vary even more on new datasets.
Nevertheless, each one of them provides a new means for investigating musical pattern discovery algorithms, namely creating a polling curve from the combination of algorithms (\PP{}), performing a comparative classification (\CC{}), and using synthetic data to test the hypotheses about the algorithms. 

\textbf{Future Work}
In future work, this research could be extended in several directions.
Firstly, the coverage of the algorithms could be extended.
We did not perform a comprehensive comparison of all musical pattern discovery algorithms given limited source code availability and data format compatibility; we also used the default configurations for each algorithm due to the large space of parameter values that would need to be explored.
Secondly, for synthetic data, we could add a greater variety of structures, possibly nested, during the data synthesis process.
Adding rhythmic variations and polyphonic patterns would make our methodology suitable to investigate algorithms that consider these aspects.
Lastly, a natural next step would be to use our findings to improve pattern discovery algorithm design.
This could be difficult, however, because the concrete steps needed to improve an algorithm is contingent on its inner workings, which may differ from algorithm to algorithm. 
In addition, there is often not a single way to improve a given algorithm, but rather multiple strategies that can be used to address the issues we have identified.
For example, an algorithm designer might add extra filters based on our classification results (Section 3), consider rhythmic aspects to a greater extent, or provide pre-configured parameter sets optimised for different application scenarios. 
Establishing which of these strategies are best suited to improving each individual algorithm is outside the scope of this paper. 

On a more general note, a mapping between different algorithms and the application scenarios they are best suited to would be desirable, considering the diversity of pattern discovery tasks.
The disagreement we found between algorithms gives support to this claim, with the differences serving as a guide for selecting algorithms based on different applications.
For example, the algorithms that discover long and overlapping patterns might be more suitable for compression, and the algorithms that discover shorter patterns might be more suitable for music theoretical and educational purposes. 

Switching from the perspective of user to the designer, pattern discovery algorithm designers also benefit from having the application scenarios as clear goals; for example, subsequent classification, segmentation, prediction, generation, and compression tasks.
Alternatively, as mentioned above, one user-friendly design would be to have different preset parameter settings for each of these different application scenarios.

Furthermore, starting from a concrete application scenario would help to clarify which classes of patterns one wishes to extract, which would, in turn, give better grounding to different evaluation methodologies. 
For example, the patterns extracted for a compression task might be very dissimilar to patterns extracted for generation purposes.
The evaluation of pattern discovery algorithms can, therefore, be diverse and go beyond the manual annotations of a single human expert as the ground truth.

One crucial link for diversifying the evaluation step is to bridge the gap between the application scenario and how we formulate the strategies used in our data synthesis approach.
With careful design, the data synthesis approach can potentially be tuned to simulate different application contexts.
Application-driven thinking combined with the data synthesis approach also has the potential to save time when annotating the data and address issues pertaining to ambiguity in the annotated patterns \cite{ren2018investigating}.
Another approach is to gather annotations from more than one person, which has been shown to be effective in \cite{koops2019annotator}. 
Gathering more annotations is a time-consuming and expensive task, requiring the participation of many different parties.
There have nevertheless been ongoing efforts in this regard.
Collecting and comparing patterns from multiple human annotators for the same piece will allow to study listeners' agreement or disagreement on patterns and finding ways to explore different degrees of consensus between the annotations.  Studies on annotator agreement on pattern discovery have the potential to enable ground truth data for pattern discovery algorithms that have a greater validity for evaluating pattern discovery algorithms than the currently used annotation data sets.

Musical pattern discovery is an interdisciplinary area of research.
Over the years, we have seen exuberant interest and considerable advancement of the state-of-the-art from contributors with diverse backgrounds. 
This has brought unique challenges as well as opportunities with a diverse range of tools, formulations, and evaluation methods, including but not limited to \PP{}, \CC{}, and the use of synthetic data with planted patterns in this paper.
We expect that greater coherence will be created by the feedback loop between such multifaceted evaluation results and the algorithms performing on diverse types of data. 
More concretely, users with access to a greater range of algorithm evaluation methodologies will be better equipped to provide feedback to algorithm designers that will ultimately be used for the design of those algorithms.

\section*{Acknowledgement}
We heartily thank the authors of the algorithms for providing their algorithms
and/or extracted pattern data and sharing their insights; the discussion with colleagues in the Department of Information and Computing Sciences, and the Centre for Complex Systems Studies, Utrecht University. 

\printbibliography

\end{document}